%% file: paper.tex
\documentclass[10pt,journal,compsoc]{IEEEtran}

\usepackage{graphicx}
\usepackage{subfig}
\usepackage{makecell}
\usepackage{tabularx}
\usepackage{balance}
\usepackage{ragged2e}
\usepackage{xcolor}
\usepackage{amsmath}
\usepackage{url}

\ifCLASSOPTIONcompsoc
  \usepackage[nocompress]{cite}
\else
  \usepackage{cite}
\fi

\newcommand{\name}{{BLOCKBENCH}}
\begin{document}
%
\title{Untangling Blockchain: A Data Processing View of Blockchain Systems}

\author{Tien~Tuan~Anh~Dinh,~\IEEEmembership{}
	    Rui~Liu,~\IEEEmembership{}
        Meihui~Zhang,~\IEEEmembership{Member, ~IEEE,}
        Gang~Chen,~\IEEEmembership{Member,~IEEE,}
        Beng~Chin~Ooi,~\IEEEmembership{Fellow,~IEEE,}
        and~Ji~Wang~\IEEEmembership{}
\IEEEcompsocitemizethanks{
\IEEEcompsocthanksitem Tien~Tuan~Anh~Dinh, Rui~Liu, Ji~Wang, Beng~Chin~Ooi are with the Department
of Computer Science, National University of Singapore, Singapore.\protect\\
E-mail: \{dinhtta, liur, wangji, ooibc\}@comp.nus.edu.sg
\IEEEcompsocthanksitem Meihui~Zhang is with Singapore University of Technology and Design, Singapore.\protect\\
E-mail: meihui\_zhang@sutd.edu.sg
\IEEEcompsocthanksitem Gang~Chen is with Zhejiang University.\protect\\
E-mail: cg@zju.edu.cn}
}



\IEEEtitleabstractindextext{%
\begin{abstract}
Blockchain technologies are gaining massive momentum in the last few years. Blockchains are distributed
ledgers that enable parties who do not fully trust each other to maintain a set of global states. The parties
agree on the existence, values and histories of the states. As the technology landscape is expanding rapidly,
it is both important and challenging to have a firm grasp of what the core technologies have to offer,
especially with respect to their data processing capabilities. In this paper, we first survey the state of the art,
focusing on private blockchains (in which parties are authenticated). We analyze both in-production and
research systems in four dimensions: distributed ledger, cryptography, consensus protocol and
smart contract. We then present \name, a benchmarking framework for understanding performance of private
blockchains against data processing workloads. We conduct a comprehensive evaluation of three major
blockchain systems based on \name, namely Ethereum, Parity and Hyperledger Fabric. The results demonstrate
several trade-offs in the design space, as well as big performance gaps between blockchain and database
systems.  Drawing from design principles of database systems, we discuss several research directions for
bringing blockchain performance closer to the realm of databases. 
\end{abstract}

}

\maketitle

\IEEEdisplaynontitleabstractindextext

\IEEEpeerreviewmaketitle

\input{intro}

\input{background}

\input{concept}

\input{tax}
\input{bb}
\input{benchmark}

\input{db}

\input{conclusion}





\section*{Acknowledgements}
This work is funded by the National Research Foundation, Prime Minister’s Office, Singapore, under its
Competitive Research Programme (CRP Award No. NRF-CRP8-2011-08). We would like to thank colleagues who have
provided valuable feedback to help improve the paper. 

\ifCLASSOPTIONcaptionsoff
  \newpage
\fi

\bibliographystyle{IEEEtran}
\bibliography{ref}

\end{document}

%% file: intro.tex
\section{Introduction}
\label{sect:intro}
Blockchain technologies are taking the world by storm, largely due to the success of Bitcoin~\cite{nakamoto2008bitcoin}.
A blockchain, also called distributed ledger, is essentially an append-only data structure maintained by a set of nodes
which do not fully trust each other. Nodes in the blockchain agree on an ordered set of blocks, each
containing multiple transactions, thus the blockchain can be viewed as a log of ordered transactions. In the database
context, blockchain can be viewed as a solution to distributed transaction management: nodes keep replicas of the data
and agree on an execution order of transactions.  However, traditional databases assume a trusted environment and employ
well known concurrency control techniques~\cite{qian16,thomson12,bailis14} to order transactions.  Blockchain's key
property is that it assumes nodes behave in arbitrary (or Byzantine) manner. Being able to tolerate Byzantine
failure by design, blockchain offers stronger security than incumbent database systems.

In the original design, Bitcoin's blockchain stores {\em coins} as the system states. For
this application, Bitcoin nodes implement a simple replicated state machine model which moves coins from one
address to another. Since then, blockchain has grown beyond crypto-currencies to support user-defined states and Turing
complete state machine models. For example, Ethereum~\cite{ethereum} enables any decentralized,
replicated applications known as {\em smart contracts}. More importantly, interest from the industry has started to
drive development of new blockchain platforms designed for private settings where participants are
authenticated.  Blockchain systems in such environments are called private (or {\em permissioned}), as opposed to the
early systems operating in public environments (or {\em permissionless}) where anyone can join and leave. Applications
such as security trading and settlement~\cite{ripple}, asset and finance management~\cite{melonport,morgan16}, banking
and insurance~\cite{gs16} are being built and evaluated. These applications are currently supported by enterprise-grade
database systems like Oracle and MySQL, but blockchain has the potential to disrupt this status quo because it incurs
lower infrastructure and human costs~\cite{gs16}. { In particular, blockchain's immutability and transparency help
reduce human errors and the need for manual intervention due to conflicting data. Blockchain can help streamline
business processes by removing duplicate efforts in data governance.} Goldman Sachs estimated 6 billion saving in
current capital market~\cite{gs16}, and J.P.  Morgan forecast that blockchains will start to replace currently redundant
infrastructure by 2020~\cite{morgan16}.     

\vspace{0.2cm}
Amid the growing commercial and academic interest, a large number of blockchain systems have sprung up, each claiming
some unique capabilities. Both private and public sector are clamoring to adopt blockchains, but they face overwhelming
choices. While challenging, it is important to have a firm grasp on what the technology can and cannot do. A
quest for understanding blockchain must ultimately answer the following questions: 
\begin{enumerate} 
  \item What is a blockchain? Specifically, what are its unique properties that benefit current and future applications?
	\item How do current blockchains differ from each other, both qualitatively in the design and quantitatively in their
  performance?
	\item What are the current challenges? And what do future blockchains look like?
\end{enumerate}
To answer these questions, in this paper, we start by distinguishing two major classes of blockchain systems,
namely public and private blockchains. We then explain four key technical concepts by which current systems
can be categorized: distributed ledger, cryptography, consensus protocol and smart contract. Next, we
describe \name~\cite{blockbench-sigmod}, our benchmarking framework for quantitatively evaluate and compare
private blockchains. Using \name, we conduct comprehensive evaluation of three major blockchains:
Ethereum~\cite{ethereum}, Parity~\cite{parity} and Hyperledger~\cite{hyperledger}. The results show that
current blockchains' performance is limited, far below what a state-of-the-art database system can offer.
Finally, we draw from our experience in building large-scale database systems several design principles that
can improve future blockchains.

In summary, our contributions are:
\begin{enumerate}
\item We provide an in-depth survey of blockchain systems. We discuss state of the art, and categorize current
systems along four dimensions: distributed ledger, cryptography, consensus protocol and smart contract.   

\item We describe our benchmarking framework, \name, that is designed for understanding performance of private
blockchains against data processing workloads. 

\item We present a comprehensive evaluation of Ethereum, Parity and Hyperledger. The results show the
limitation of blockchains as data processing platforms. They identify several performance bottlenecks, and therefore can
serve as a baseline for future blockchain research and development.  \end{enumerate} 

In the next section, we provide an
overview of blockchain systems, separating them into public and private
settings. Section~\ref{sec:concept} explains the four building blocks which are used in Section~\ref{sec:tax}
to categorize existing blockchains. Section~\ref{sec:blockbench} describes \name, followed by the evaluation of
three blockchains in Section~\ref{sec:evaluation}. Section~\ref{sec:db} discusses a number of lessons
learned from the performance study, and how to bring design principles from databases to improve
blockchains. Section~\ref{sec:conclusion} concludes.

%% file: background.tex
\section{Blockchains: Private vs. Public}
\label{sec:background}


\begin{figure}
\centering
\includegraphics[width=0.48\textwidth]{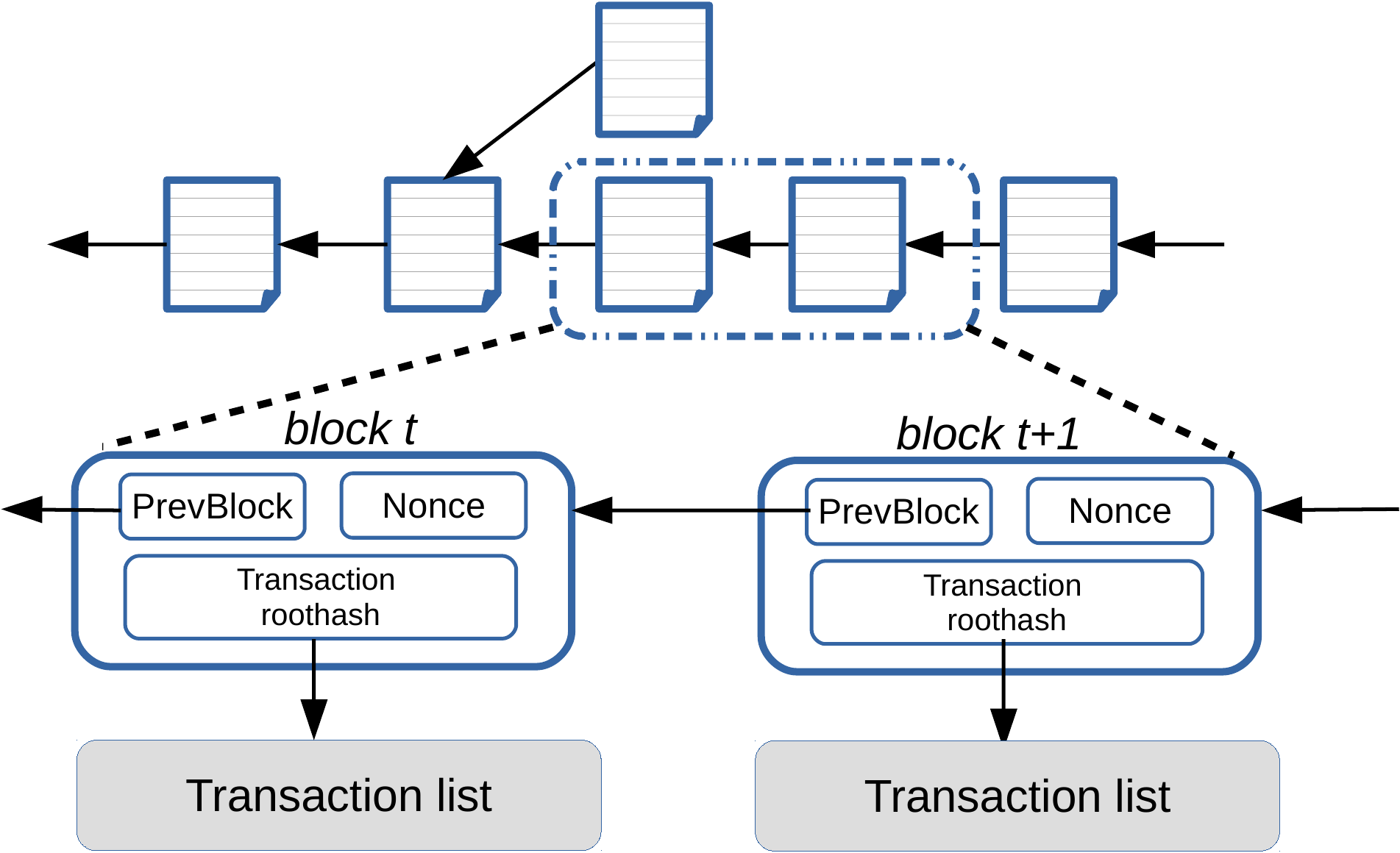}
\caption{Blockchain data structure. Transactions are packed into blocks which are linked to previous blocks.}
\label{fig:stackbc}
\end{figure}

A typical blockchain system consists of multiple nodes which do not fully trust each other. Some nodes exhibit Byzantine
behavior, but the majority is honest. Together, the nodes maintain a set of shared, global states and perform
transactions modifying the states. Blockchain is a special data structure which stores historical states and
transactions.  All nodes in the system agree on the transactions and their order.  Figure~\ref{fig:stackbc} shows the
blockchain data structure, in which each block is linked to its predecessor via a cryptographic pointer, all the way
back to the first (genesis) block. Because of this, blockchain is often referred to as a distributed ledger. 

A transaction in a blockchain is the same as in traditional databases: a sequence of operations applied on some states.
As such, the blockchain transaction requires the same ACID semantics. The key difference is the failure model under
consideration. Current transactional, distributed databases~\cite{hstore, spanner} employ classic concurrency control
techniques such as two-phase commit to ensure ACID.  They can achieve high performance, because of the simple crash failure
model.  In contrast, the original blockchain design considers a much hostile environment in which
nodes exhibit Byzantine behavior. Under this model the overhead of concurrency control is much
higher~\cite{castro1999practical}.  

At a high level, a blockchain system can be categorized as either public or private. In the former, any node can join
and leave the system, thus the blockchain is fully decentralized, resembling a peer-to-peer system~\cite{p2p}. In the
latter, the blockchain enforces strict membership. More specifically, there is an access control mechanism to determine
who can join the system. As the result, every node is authenticated and its identity is known to the other nodes.

\subsection{Public Blockchain}
Bitcoin~\cite{nakamoto2008bitcoin} is the most well known example of public blockchains.  In Bitcoin the states are
digital coins (crypto-currencies), and a transaction moves coins from one set of addresses to another.
Each node broadcasts a set of transactions it wants to perform.  Special nodes called {\em miners} collect transactions
into blocks, check for their validity, and start a consensus protocol to append the blocks onto the blockchain. Bitcoin
uses {\em proof-of-work} (PoW) for consensus: only a miner which has successfully solved a computationally hard puzzle
(finding the right nonce for the block header) can append to the blockchain. PoW is tolerant of Byzantine failure, but
it is probabilistic in nature: it is possible that two blocks are appended at the same time, creating a {\em fork} in
the blockchain. Bitcoin resolves this by only considering a block as confirmed after it is followed by a number of
blocks (typically six blocks). This probabilistic guarantee leads to security and performance issues: attacks have
been demonstrated by an adversary controlling only 25\% of the nodes~\cite{eyal14}, and Bitcoin transaction throughput
remains very low (7 transactions per second~\cite{croman2016scaling}). 

Most public blockchain systems employ variants of PoW for consensus. PoW works well in the public settings because it
guards against Sybil attacks~\cite{p2p}. However, being non-deterministic and computationally expensive, it is
unsuitable for applications such as banking and finance which must handle large volumes of transactions in a deterministic
manner.  

\subsection{Private Blockchain}
Hyperledger~\cite{hyperledger} is among the most popular private blockchains. Since node identities are known in the
private settings, most blockchains adopt one of the protocols from the vast literature on distributed consensus.
Zab~\cite{zab}, Raft~\cite{raft}, Paxos~\cite{paxos}, PBFT~\cite{castro1999practical} are popular protocols that are in
active use today. Hyperledger directly uses PBFT\footnote{Hyperledger has two main releases: v0.6.0 and v1.0.0-rc1. The
former supports PBFT, but the latter adopts a no-Byzantine consensus protocol based on Kafka.}, while others like
Parity\cite{parity}, Ripple~\cite{ripple} and ErisDB~\cite{monax} develop their own variants. PBFT is a
three-phase protocol. In the {\em pre-prepare} phase, a leader broadcast a value to be commit by other nodes.
Next, in the {\em prepare} phase, the nodes broadcast the values they are about to commit. Finally, the {\em
commit} phase confirms the committed value when more than two third of the nodes agree in the previous phase.
PBFT is communication bound, but it achieves both safety and liveness in partially synchronous networks.  Besides
deterministic consensus, another key property of private blockchains is that they support smart contracts which can
express highly complex transaction logics. These properties are particularly desirable in business and financial systems.
Indeed, private blockchains evoke such interest from major banking and financial institutions that some even claim that
they have the potentials to disrupt current practices in data management~\cite{gs16,morgan16}.





%% file: concept.tex
\section{Key Concepts}
\label{sec:concept}
Categorizing blockchains as public or private is useful for identifying major characteristics of many
blockchains. However, understanding their subtle differences warrants a finer taxonomy.  This section introduces four
underpinning concepts, based on which a more detailed classification of the systems can be obtained. 

%

\subsection{Distributed Ledger}
\begin{table*}
	\caption{Examples of distributed ledgers.}
	\label{table:cmpdl}
	\centering  
	\begin{tabular}{m{2.5cm} | m{2.5cm} | m{2.5cm} | m{6cm}}
		\Xhline{1.2pt}
		\hline
		\hline
	  \bf{Data Model} & \bf{Number of ledgers} & \bf{Owner} & \bf{Example}\\
		\hline
	     Accounts & One & Administrator & Traditional ledgers used in financial institutions.\\
		\hline 
		Assets & Many & Group of users & Private ledger used within a financial institution, or between small groups of
    financial organizations, e.g. global financial services.\\
		\hline 
		Coins or accounts & One & Any user & Crypto-currencies like Bitcoin or Ethereum. \\
		\hline
    \hline
		\Xhline{1.2pt}
	\end{tabular}
\end{table*}

A ledger is a data structure that consists of an ordered list of transactions. For example, a
ledger may record monetary transactions between multiple banks, or goods exchanged among known parties. In
blockchains, the ledger is replicated over all the nodes. Furthermore, transactions are grouped into blocks which are
then chained together. Thus, the distributed ledger is essentially a replicated append-only data structure. A blockchain
starts with some initial states, and the ledger records entire history of update operations made to the states.

A system supporting distributed ledgers can be characterized in three dimensions, as illustrated in
Table~\ref{table:cmpdl}. First, the application built on top of the ledger determines the data model of what being
stored in the ledger. For example, a crypto-currency application may adopt the user-account model resembling traditional
banking systems. On the other hand, a general-purpose blockchain may use low-level model such as table or
key-value. Second, the system may have one or more ledgers which may be connected to each other. A large enterprise, for
example, may own multiple ledgers, one for each of its departments: engineering, customer care, supply chain, payroll,
etc.. Third, ledger ownership may vary from completely open to public to strictly controlled by one party. Bitcoin, for
example is completely open, and as a consequence requires expensive consensus protocol to identify who can update the
ledger. Parity~\cite{parity}, on the other hand, pre-determines a set of owners who can write to the ledger simply by
singing the blocks.


\subsection{Consensus}
The content of the ledger reflects historical and current states maintained by the blockchain. Being
replicated, updates to the ledger must be agreed on by all parties. In other words, multiple parties must come
to a consensus. Note that this is not the case in many real-world applications such as fiat currency, in which
one entity (e.g. the bank or the government) decides the updates.

One key property of a blockchain system is that the nodes do not trust each other, meaning that some may
behave in Byzantine manners. The consensus protocol must therefore tolerate Byzantine failures.  The research
literature on distributed consensus is vast, and there are many variants of previously proposed protocols
being developed for blockchains~\cite{vulkolic15}. They can be largely classified along a spectrum. One
extreme consists of purely computation based protocols that use proof of computation to randomly select a
node which single-handedly decides the next operation.  Bitcoin's proof-of-work (PoW) is an example. The other
extreme are purely communication based protocols in which nodes have equal votes and go through multiple
rounds of communication to reach consensus. These protocols, PBFT~\cite{castro1999practical} being the prime
example, are used in private settings because they assume authenticated nodes. In between these extremes are
hybrid protocols which aim to improve performance of PoW and PBFT.  For instance, Proof-of-Elapsed-Time (PoET)
eliminates expensive mining in PoW by leveraging trusted hardware such as Intel SGX. Another example is
Proof-of-Authority (PoA)~\cite{poa} which improves PBFT by pre-selecting a small set of trusted nodes that
vote among themselves to reach consensus. Similarly, Stellar~\cite{stellar} and Ripple~\cite{ripple} improve
PBFT by executing consensus in smaller networks.




\subsection{Cryptography}
Blockchain systems make heavy use of cryptographic techniques to ensure integrity of the ledgers.  Integrity
here refers to the ability to detect tampering of the blockchain data. This property is vital in public
settings where there is no pre-established trust. For example, public confidence in crypto-currencies like
Bitcoin, which determines values of the currencies, is predicated upon the integrity of the ledger; that is
the ledger must be able to detect double spending. Even in private blockchains, integrity is equally essential
because the authenticated nodes can still act maliciously.  

There are at least two levels of integrity protection. First, the global states are protected by a hash (Merkle)
tree whose root hash is stored in a block. Any state change results in a new root hash.  The tree's
leaves contain the states, the internal nodes contain the hashes of their children. For instance, Hyperledger v0.6 uses
a bucket hash tree, in which states are grouped (by hashing) into a pre-defined number of buckets.  Ethereum, on the other
hand, employs a Patricia-Merkle tree which resembles a trie and whose leaves are key-value states. Second, the
block history is protected, that is the blocks are immutable once they are appended to the blockchain. The key
technique is to link the blocks through a chain of cryptographic hash pointers: the content of block number
$n+1$ contains the hash of block number $n$. This way, any modification in block $n$ immediately invalidates
all the subsequent blocks. By combining Merkle tree and hash pointers, blockchain offers a secure and
efficient data model that tracks all historical changes made to the global states. 

Blockchain's security model assumes the availability of public key cryptography. Identities, including user
and transaction identities, are derived from public key certificates. Secure key management, therefore, is
essential to any blockchains. As in other security systems, losing private keys means losing access.  But in
blockchain applications such as crypto-currencies, losing the keys has direct and irrevocable financial
impact.  We discuss in Section~\ref{subsec:crypto} different schemes for key and identity management.  

There exist many research systems that extend the original blockchain design with novel and complex
cryptographic protocols. They aim to improve security and performance with esoteric techniques such as
zero-knowledge proofs, group signatures and trusted hardware. We discuss them in greater detail in
Section~\ref{subsec:crypto}.

\subsection{Smart Contracts}
A smart contract refers to the computation executed when a transaction is performed. It can be
regarded as a stored procedure invoked upon a transaction. The inputs, outputs and states affected by the
smart contract execution are agreed on by every node.

All blockchains have built-in smart contracts that implement their transaction logics. In crypto-currencies,
for example, the built-in smart contract first verifies transaction inputs by checking their signatures. Next,
  it verifies that the balance of the output addresses matches that of the inputs. Finally, it applies changes
  to the states. In the rest of the paper we do not refer to such built-in logics as smart contracts. Instead,
  we only consider smart contracts that can be defined by users.

One way to characterize a smart contract system is by its language. At one extreme, Bitcoin provides fewer
than 200 {\em opcodes} from which users can write stack-based scripts. For example, the following script
verifies if 2 out of 3 valid signatures are available.  

{\small
\begin{verbatim}
OP_2 <Pub1> <Pub2> <Pub3> OP_3 OP_CHECKMULTSIG
\end{verbatim}
}
\noindent At the other extreme, Ethereum smart contracts can specify arbitrary computations, i.e. they are Turing
complete code.  Figure~\ref{fig:doubler} shows a snippet of a real smart contract running on Ethereum. It
implements a pyramid scheme: users send money to this contract which then pays interests to early
participants.  The contract has its own states, namely the list of participants, and exports a function called
{\tt enter}. A user invokes the contract by sending his money through a transaction. When executed, the
contract can access the input address (user account) via {\tt msg.sender} and the transaction value via {\tt
msg.amount}. It updates the accumulated balance, computes the interest for each participants. Finally, payment
is made by invoking {\tt etherAddress.send.}

\begin{figure}
	\centering
	{\footnotesize
		\begin{verbatim}
		contract Doubler {
		  struct Partitipant {
		    address etherAddress;
		    uint amount; 
		  }
		  Partitipant[] public participants;
		  uint public balance = 0; 
		  ...
		  function enter() {
		    ...
		    balance+= msg.value;
		    ...
		    if (balance > 
              2*participants[payoutIdx].amount) {
		      transactionAmount = ...
		      participants[payoutIdx].
              etherAddress.send(transactionAmount); 
		      ...
		    }
		  }
		  ...
		}
		\end{verbatim}
	}
	\caption{An example of Ethereum smart contract, written in Solidity, which implements a pyramid scheme.}
	\label{fig:doubler}
\end{figure}

In between the two extremes are smart contract systems that offer more expressiveness than Bitcoin's opcodes,
but they reject Turing-completeness. Kadena~\cite{kadena} and BigchainDB~\cite{bigchaindb} support contracts
with complex, but constrained semantics so that they can be formally checked for safety. 

\vspace{0.7em}
Another way to categorize smart contract systems is by their runtime environments. Most systems execute smart
contracts in the same runtime as the rest of the blockchain stack. We refer to them as employing
native runtimes. For example, Kadena parses contracts written in its Haskell-like language and executes them directly
as Haskell programs.  Ethereum, on the other hand, comes with its own virtual machine for executing Ethereum
bytecodes. Hyperledger, opting for portability, employs Docker containers to execute its contracts.

%% file: tax.tex
\section{State of the Art}
\label{sec:tax}
\begin{table*}
	\caption{Comparison of blockchain systems. Ones in italics are deemed inactive or at early phases  
  of development.}
	\label{table:statb}
	\centering  
	\begin{tabular}{m{1.8cm}<{\centering} | m{2.3cm}<{\centering} | m{2.3cm}<{\centering} | m{3.5cm}<{\centering} | m{2.1cm}<{\centering} | m{2.7cm}<{\centering}}
		\Xhline{1.2pt}
		\hline
		\hline
		& \bf{Application} & \bf{Smart contract execution} & \bf{Smart contract language} & \bf{Data model} & \bf{Consensus}\\
		\hline
		Hyperledger v0.6.0 \cite{hyperledgerv06} & General applications& Dockers & Golang, Java & Key-value& PBFT\\
		\hline
		Hyperledger v1.0.0 \cite{hyperledgerv1} & General applications & Dockers & Golang, Java & Key-value & Ordering service (Kafka)\\
    \hline
    Bitcoin & Crypto-currency & Native & Golang, C++ & Transaction-based & PoW \\
    \hline
    Litecoin~\cite{litecoin} & Crypto-currency & Native & Golang, C++ & Transaction-based & PoW (memory) \\
    \hline
    ZCash~\cite{sasson14}& Crypto-currency & Native & C++ & Transaction-based & PoW (memory) \\ 
    \hline 
		Ethereum \cite{ethereum} & General applications& EVM & Solidity, Serpent, LLL & Account-based & PoW\\
		\hline 
		Multichain \cite{multichain} & Digital assets & Native & C++ & Transaction-based& Trusted validators (round robin)\\
		\hline 
    Quorum~\cite{quorum} & General applications& EVM & Golang & Account-based & Raft \\
    \hline
		{\em HydraChain \cite{hydrachain}} & {\em General applications} & {\em Python, EVM} & {\em Solidity,
    Serpent, LLL} &
    {\em Account-based} & {\em Trusted validators (majority)}\\
		\hline 
		{\em OpenChain \cite{openchain}} & {\em Digital assets} & {-} & {-} & {\em Transaction-based} &
    {\em Single validator}\\
		\hline 
		{\em IOTA \cite{iota}} & {\em Digital assets} & {-} & {-} & {\em Account-based} & {\em IOTA's
    Tangle Consensus}\\
		\hline 
		BigchainDB \cite{bigchaindb} & Digital assets & Native & Python, crypto-conditions & Transaction based & Trusted
    validators (majority)\\
		\hline
		Monax \cite{monax} & General applications& EVM & Solidity & Account-based &
    Tendermint~\cite{tendermint}\\
		\hline
		Ripple~\cite{ripple} & Digital assets & - & - & Account-based& Ripple consensus\\
		\hline
		{\em Kadena \cite{kadena}} & {\em Pact applications} & {\em Native} & {\em Pact} & {\em Table} &
    {\em ScalableBFT~\cite{tangaroa}}\\
		\hline
		Stellar \cite{stellar} & Digital assets & - & - & Account-based & Stellar consensus\\
		\hline
		{\em Dfinity \cite{dfinity}}& {\em General applications} & {\em EVM} & {\em Solidity, Serpent, LLL} & {\em Account-based}
    & {\em Threshold relay}\\
		\hline
		Parity \cite{parity} & General applications & EVM & Solidity, Serpent, LLL & Account-based & Trusted validators
    (round robin)\\
		\hline
		Tezos \cite{goodman2014tezos} & Michaleson applications& Native & Michaleson & Account-based & Proof of Stake\\
		\hline
		Corda \cite{brown2016introducing} & Digital assets & JVM & Kotlin, Java & Transaction-based & Raft\\
		\hline
		Sawtooth Lake \cite{sawtooth} & General applications& Native  & Python & Key-value & Proof of elapsed time\\\hline
		\hline
		\Xhline{1.2pt}
	\end{tabular}
\end{table*}

In this section we compare current blockchains using the four concepts discussed in
Section~\ref{sec:concept}. We explain their design in more detail and highlight their subtle differences. We also
discuss research problems that are being tackled. 

A list of blockchains and their properties are shown in Table~\ref{table:statb}. Major systems are included,
but we stress that the list is not exhaustive, especially given the growing commercial and academic interest in
blockchains.  Systems shown in italics are either no longer in active development, or are still in initial
phases of development. For examples, Hydrachain~\cite{hydrachain} codebase was last updated about 8 months ago
at the time of writing, and IOTA's current codebase consists of only a reference implementation\footnote{As
many other blockchain projects do, IOTA is raising fund for development via token sale.}. The table has no
column for cryptography, since all systems (except for ZCash) employ standard techniques described in
Section~\ref{sec:concept}. Novel cryptographic protocols that are not yet integrated are discussed in
Section~\ref{subsec:crypto}.

\subsection{Distributed Ledger}
Recall that a system supporting distributed ledgers is characterized by its target applications, by the number of
ledgers, and by the ledger ownership. In the following, we group the systems listed in Table~\ref{table:statb} by their
target applications. 

\subsubsection*{Crypto-currency} The most successful adoption of blockchain technology is crypto-currency. In
the wake of Bitcoin's success, multiple competing currencies appear. Most of these alternative currencies (or
alt-coins) such as Litecoin or Dodgecoin, adopt similar data models to Bitcoin's. Ethereum, departs from
Bitcoin's transaction-based model and instead implements an account-based model. The nature of currency
applications requires that the ledger must be open and the system maintains only one ledger.    

\subsubsection*{Digital assets} Crypto-currency is one instance of digital assets --- pieces of data with
attached real-world values. Unlike crypto-currencies which are created on and derive their values directly
from the blockchains, digital assets are often issued by real world entities and blockchains are merely a medium
to record their existence and exchanges.  Multichain~\cite{multichain}, BigchainDB and Corda offer ledgers for storing and
tracking asset history. Like Bitcoin, their data models are transaction-based which are centered around
assets.  These systems target private settings, in which multiple organizations can spin up a network to trade
assets among each other. The organizations are the ledger owners, and it is common to have more than one
ledger among them. Stellar, Ripple and IOTA issue their own assets (tokens) and offer their ledgers as a
medium of exchange or a platform for micro-payment. IOTA, in particular, allows for zero-fee micropayment via
its tokens, which makes the ledger useful for exchanges among IoT devices. The ledgers in these systems adopt
account-based data models. One ledger exists per system and it is open; that is anyone can buy tokens and take
part in the exchanges.

\subsubsection*{General applications} Going beyond crypto-currency and asset management, some ledgers
support running general, user-defined computations (or smart contracts). Ethereum and its derivatives, namely
Hydrachain, Quorum, Monax, Parity and Dfinity let users write arbitrary business logics executed on top of the
ledger. For example, Ethereum contracts range from simple crowdfunding campaigns to complex investment funds
like the DAO~\cite{dao}. Dfinity has a special type of contract --- the governance contract --- that enforces
real-world regulations on Ethereum-like blockchains.  Hyperledger and its close cousin Sawtooth Lake likewise
support running Turing-complete code. They offer key-value data model, with which the applications can create and
update key-value tuples on the blockchain. Kadena and Tezos restrict how powerful the applications can be by
devising their own languages which are not Turing complete but can be formally verified.  Tezos data model is
account-based, whereas Kadena's is based on table. In particular, Kadena applications operate on {\em key-row}
structures with schemas, versions and column history.

\subsection{Cryptography}
\label{subsec:crypto}
\subsubsection*{Identity management} A user in a blockchain is uniquely identified by her public key certificate.
In public settings, the user first generates a key pair (the default option being ECDSA based on the Secp256k1
elliptic curve), then derives the identity as the hash of the public key. This hash serves as a transaction
address or an account number in crypto-currencies systems. To claim ownership of the transaction output or of
the account, the user signs transactions with the corresponding private key. In private settings, there is an
additional access control layer. Hyperledger separates this layer from the blockchain, in the form of a
membership provider service and a certificate authority service. The administrator can implement arbitrary
policies with these services to control who gets access to the blockchain.  Signed requests sent to
Hyperledger are first checked against these services before processed by the next (consensus) component.
Multichain offers a simpler model with a fixed number of global permissions, while the remaining systems
provide little detail of their protocols. 

The problem of managing user keys is the same in private blockchains as in typical enterprise systems,
thus existing solutions can be readily integrated. In public blockchains, however, the sheer scale and
monetary impacts of losing private keys calls for more secure and more usable protocols. Bitcoin, in particular,
embodies the challenges, as Bitcoin users themselves are tasked with managing large numbers of keys which are
refreshed on new transactions. Eskandari et al.~\cite{eskandari15} evaluated six approaches for Bitcoin key
management (or wallet): local storage, password protected storage, offline storage, air-gapped storage,
password-derived keys, and hosted storage. The authors found that none of these approaches is satisfactorily
usable, due to misuse of metaphors to traditional currencies, and also due to confusing abstractions.  

\subsubsection*{Trusted hardware} Recent distributed systems are leveraging trusted hardware such as Intel SGX and ARM
TrustZone to improve performance at a slight cost of security~\cite{m2r, opaque}. Sawtooth Lake proposes proof of elapsed time (PoET) as an
efficient replacement for proof-of-work. TownCrier~\cite{towncrier} employs SGX to implement a trusted party for
vetting external contents and importing them to the blockchain. These systems are based on a trust model that is
weaker than that of a purely cryptographic system. In particular, their security is dependent upon a trusted
computing base (TCB) that is running inside the trusted hardware. Smaller TCBs mean better security.

All systems based on trusted hardware rely on {\em remote attestation} protocols. A key pair, called Endorsement
Key (EK), is burnt into each device during manufacturing. Such a key pair serves as the root of trust, from
which other short-term keys are derived. Before a piece of code is loaded, the hardware {\em measures} it by
hashing the code content and signing it with one of the keys. The signed measurement together with the key
certificate attest to a remote party what is being run in the local device. This protocol requires a
certificate authority that maintains and endorses a list of known certificates and a list of revoked
certificates. Highly complex attestation schemes, for example direct anonymous attestation~\cite{daa}, offer
hardware anonymity without a certificate authority.

\subsubsection*{Transaction privacy} Most blockchains are designed to protect transaction integrity, but they do
not consider transaction privacy. A blockchain is said to have transaction privacy when (1) transactions cannot be
linked from one to another, and (2) the transaction content is known only to its participants. In private
settings, complete transparency of transaction history may not be a problem. Either transparency is desirable for
the applications, such as financial auditing, or it is straightforward to add an access control layer to protect the
blockchain data. In public settings, on the other hand, the need for transaction privacy is driven by two
factors. First, deanonymization attacks have successfully recovered the underlying structure of the Bitcoin
network~\cite{meiklejohn13}, and even linked Bitcoin addresses to real-world identities~\cite{chainalysis}.
Second, transaction linkability can undermine the currency's fungibility, rendering some coins more valuable than
others due to their histories. 

Zerocoin~\cite{miers13} is the first blockchain providing transaction unlinkability. It extends Bitcoin to allow
for trading between bitcoins and special coins called {\em zerocoins}. Zerocoin essentially implements a cryptographic mixer that
hides linkability between zerocoins and the corresponding bitcoins. Each zerocoin is a cryptographic commitment to two
random values $(s,r)$. When redeeming a zerocoin, the owner reveals $s$ as the proof that the coin has not been spent, and a
zero-knowledge proof of $r$. Transaction unlinkability is derived from the fact that the coin being
redeemed can be any of the many unspent zercoins. 

Zerocash~\cite{sasson14} extends Zerocoin by improving efficiency of the latter's cryptographic operations. It functions as
a stand-alone blockchain, as opposed to Zercoin being an extension of Bitcoin. Transactions in Zerocash, including split
and merge transactions, are fully private. They are based on complex zero-knowledge proofs which only reveal the fact
that there exists unredeemed coins whose sum is a specific value. Both Zerocoin and Zerocash are implemented using
zkSNARK~\cite{zksnark} and carry large overheads due to the underlying zero-knowledge protocols. Zerocash, for instance,
requires a trusted party to securely create and distribute some public parameters whose sizes are in hundreds of
megabytes.  

\subsubsection*{Advanced signatures} Bitcoin supports multi-signatures, in which a transaction can be redeemed
when at least $t$ out of $n$ valid signatures are available. Multi-signatures are resilient against corrupt
individuals by virtue of spreading the decryption or singing capabilities to a group of users. Lightning
Network~\cite{lightning}, an extension of Bitcoin with near-instant payment confirmation, relies on
multi-signatures to first deposit some mutual funds on the blockchain.  Once confirmed, payments from the funds can
happen outside of the blockchain with immediate confirmation.  Finally, the funds can be closed with
corresponding transactions signed with all the required signatures.  Extensions of Bitcoin multi-signature
scheme can be built directly on top of ECDSA~\cite{goldfeder16}. More advanced schemes, e.g. ~\cite{shoup00,
bls}, can be employed (albeit not without major changes in the current
design).

Byzcoin~\cite{byzcoin} uses a group signature scheme called Cosi~\cite{cosi} to reduce communication overhead in PBFT.
Cosi involves four rounds of communication at the end of which a collective signature is generated and verified by all
members of the group. The signature is structured as a tree of Schnorr signatures. It significantly reduces
the size of messages broadcast in the network during the prepare and commit phase of PBFT, because each node no
longer needs individual signatures from all the other for verification.

\subsection{Consensus}
\begin{table*}
	\caption{Comparison of consensus protocols.}
	\label{table:cmpcp}
	\centering  
	\begin{tabular}{m{4.6cm}<{\centering} | m{2.5cm}<{\centering} | m{8.2cm}}
		\Xhline{1.2pt}
		\hline
		\hline
	  \bf{Consensus Protocol} & \bf{Network Settings} & \bf{Description}\\
		\hline
	     PBFT-based & Private & Hyperledger uses the original PBFT~\cite{castro1999practical}.
       Tendermint~\cite{tendermint} enhances it by assigning unequal weights to votes. Other variants include
       Scalable BFT \cite{BehlDK14}, Parallel BFT \cite{zbierski2015parallel}, Optimistic BFT
       \cite{Zhao:2016:OBF:2996052.2996056}, etc. \\ 
    \hline
      Stellar & Federated & Stellar network~\cite{mazieres2015stellar} proposes its own consensus protocol
      where the nodes form intersecting groups (federates). Consensus is agreed in each group, then propagated
      to the rest of the network.\\
    \hline
      Ripple & Federated & Ripple payment system~\cite{ripple} proposes a variant of PBFT where the nodes
      belong to intersecting groups, and in each group there is a large majority of non-Byzantine nodes. \\
      \hline
	     Proof-of-Work (PoW) & Public & Bitcoin uses pure proof-of-work, which leads to scalability issues.
       Bitcoin-NG~\cite{bitcoin-ng}, Byzcoin~\cite{byzcoin} separate leader election from transaction
       validation in PoW, thus increase the overall performance. \\
      \hline
       Proof-of-Stake (PoS) & Public & Tendermint~\cite{tendermint} uses PoS, in which a node's ability to
       create new block is determined by its stake in the blockchain, e.g. the amount of currencies it
       owns~\cite{pos}. A set of high-stake owners uses another consensus mechanism, which is usually faster
       than PoW, to reach agreement on a new block.  \\ 
      \hline
      Threshold Relay & Public & Dfinity~\cite{dfinity} proposes {\em threshold relay} in which nodes form random group
      based on a public verifiable random function (Byzcoin~\cite{byzcoin} and Elastico~\cite{elastico} adopt
      similar approaches). The nodes in the group create a new block by signing it
      using threshold signature.\\
      \hline

        Proof-of-Authority (PoA) & Private & Parity~\cite{parity} uses PoA, in which some pre-defined nodes are considered
        trusted authorities and they can propose the next blocks. It then uses round-robin
        scheduling to assign every authority node a time window during which it can propose blocks.\\ 
      \hline
       Proof-of-Burn (PoB) & Public & Slimcoin~\cite{pob} uses PoB, in which a node destroys some base currencies it owns in another blockchain in
       order to get a chance of proposing a new block. Slimcoin supports PoB based on
       Peercoin~\cite{peercoin}. \\ 
       \hline
       Proof-of-Elapsed Time (PoET) & Private & Sawtooth~\cite{sawtooth} uses PoET, in which each node runs a
       trusted hardware, for example Intel SGX~\cite{poet}, that generates random timers. The first node whose
       timer has expired  can propose the next block. \\ 
	     \hline
       Others & Public & Other protocols based on PoW are of the form {\em proof-of-X}, for examples:
       Proof-of-Activity \cite{BentovLMR14}, Proof-of-Space \cite{ParkPAFG15}, Proof-of-Luck
       \cite{MilutinovicHWK17}, etc. \\ \hline
		\hline
		\Xhline{1.2pt}
	\end{tabular}
\end{table*}

Recall that there is a spectrum of consensus protocols behind blockchain systems, starting from purely
computation bound like PoW to purely communication bound like PBFT. Table~\ref{table:cmpcp} summarizes key
properties of the major protocols which we now explore in detail.  

\subsubsection*{Proof of Work variants}
All PoW protocols require miners to find solutions to cryptographic puzzles based on cryptographic hashes. 
Specifically, the solution is a random nonce $n$ such that: 

\[ 
H(n \| H(b)) < t
\]
for a cryptographic hash function $H$, a threshold $t$ and the current block content $b$. The original
  protocol implemented in Bitcoin uses SHA-256 as the hash function. The availability of custom hardware
  (ASIC) that speeds up hash computation prompts other crypto-currencies to adopt memory-hard hash functions.
  Ethereum uses Dagger-Hashimoto function, Litecoin and Dodgecoin use scrypt, and ZCash
  uses Equihash function.  These functions are resistant to ASIC, as they demand large investment in memory,
  but are easy to verify.

How fast a block is created is dependent on how hard the puzzle is. Bitcoin sets $t$ to a value equivalent of $10$
minutes per block. Litecoin, Dodgecoin and ZCash decrease $t$ to achieve lower average block time to several minutes. $t$
cannot be arbitrary small because it leads to unnecessary forks in the blockchain. Forks not only lead to
wastage of resources but have security implication since they make it possible to double spend. Ethereum adopts
GHOST~\cite{ghost} protocol which helps bring down block generation time to tens of seconds without compromising much
security. In GHOST, the blockchain is allowed to have branches as long as the branches do not contain conflicting
transactions. 

\subsubsection*{Proof of Stake}
PoW mining is hugely expensive. The process is particularly energy intensive, and has been estimated to consume enough
electricity to power a small country like Denmark~\cite{denmark}. PoS is proposed to substantially reduce the cost of mining.
Unlike Ethereum's GHOST, PoS maintains a single branch but changes the puzzle's difficulty to be inversely proportional
to the miner's stake in the network. A stake is essentially a locked account with a certain balance representing the
miner's commitment to keep the network healthy. Let $s$ be the function that returns the stake, then a miner $M$ can
generate a new block by solving the puzzle of the following form: 

\[
H(n \| H(b)) < s(M).t
\]
It can be seen that the greater the stake $s(M)$, the easier it is to find $n$. 

Peercoin, forked from Bitcoin, is among the first systems with PoS. It bootstraps by running PoW to generate
coins. The function $s(.)$ in Peercoin takes a coin $C$ as input and returns $C.age(C)$ where $age(C)$ is
the coin's age. Nxt~\cite{nxt}, another PoS system, bootstraps by selling its tokens. The function $s(.)$ in Nxt
considers both the miner's balance and the elapsed time from the last block. The longer it is since the last block,
the easy it is to solve the puzzle. In particular:

\[
s(M, b_h) = bal(M).age(b_{h-1})
\]
where $b_h$ is the current block at height $h$, $bal(M)$ returns how many coins in $M$'s account, and $age$ returns how
much time has passed since the creation of a block at a certain height. 

Ethereum's upcoming PoS protocol is implemented as a smart contract. Referred to as Casper, it allows miners to become 
{\em validators} by depositing Ethers to the Casper account. The contract then picks a validator to propose the next
block according to the deposit amount. Its unique feature, however, is to force validators to behave correctly or else
risk losing the entire deposit. In particular, each validator places a bet on whether a certain block will be confirmed
in the future. If the block is confirmed, the validator gets a small reward. But if it is not, the validator loses its
deposit. This mechanism avoids the {\em nothing-at-stake} problem in which validators can propose blocks in different
branches. Tezos implements a simplified version of Casper in which the nodes {\em buy in} to become authorities which
can then approve changes to the underlying blockchain. Tezos aims to provide an amendable blockchain in which soft forks
and hard forks are inherent features of the blockchain.  

\subsubsection*{PBFT variants}
PoW suffers from non-finality, that is a block appended to a blockchain is not confirmed until it is extended by many
other blocks. Even then, its existence in the blockchain is only probabilistic. For example, eclipse attacks on
Bitcoin~\cite{eclipse} exploit this probabilistic guarantee to allow double spending. In contrast, the original PBFT
protocol~\cite{castro1999practical} is deterministic. Implemented in the earlier version of Hyperledger (v0.6), the
protocol ensures that once a block is appended, it is final and cannot be replaced or modified. It incurs $O(N^2)$
network messages for each round of agreement where $N$ is the number of nodes in the network. In practice, however, the
original protocol scales poorly and collapses even before reaching the network limit~\cite{clement09}. We observe the
same scalability issues in our evaluation of Hyperledger with \name.  

Tendermint proposes a small modification on top of PBFT. Instead of each node having an equal vote, in Tendermint each
node may have different voting power, proportional to their stake in the network. To reach agreement in Tendermint it is
necessary to only gather over $2/3$ of the total voting power. This may be cheaper than waiting for $2/3$ of the network to
response when there is a small number of nodes with high stakes. 

Recent works on improving PBFT have mainly focused on its performance. Zyzzyva~\cite{speculative-bft} optimizes for
normal cases (when there are no failures) via speculative execution. XFT~\cite{xft}, assumes a network less hostile than
purely Byzantine, and demonstrates better performance by reducing the number of network messages.
HoneyBadger~\cite{honeybadger}, on the other hand, focuses on improving security under asynchronous networks.
It employs a {\em randomized agreement} protocol which achieves safety with overwhelming probability even
under network asynchrony. By optimizing the network layer, it is shown to outperform PBFT even when the
network is synchronous. Both Zyzzyva, XFT and HoneyBadger hold great promise, but they have not been
integrated into any blockchains. 

\subsubsection*{Trusted hardware}
Most overheads of PoW and PBFT can be attributed to the assumption that nodes behave in Byzantine manners. The
availability of Intel SGX~\cite{sgx} or ARM TrustZone~\cite{trustzone}, however, makes it possible to relax
the trust model in the Byzantine settings. In particular, a node equipped with trusted hardware can be
reliably checked for certain properties, for example, that it is running a specific software. 

Sawtooth Lake leverages SGX to replace PoW with a more efficient protocol called PoET. Specifically, PoET runs inside an
enclave protected by SGX. It starts by taking a block number as input and generating a timer of a random
duration $t$.  Afterward, it can produce certificates indicating how much time has passed since the timer starts. A
node whose PoET generates the smallest $t$ can append the block when the timer expires. In particular, the
node attaches its PoET certificate to the block, and as long as $t$ is smaller than what generated by any
other node the block is accepted. 

A2M~\cite{a2m} and Hybster~\cite{hybster} both exploit trusted hardware to reduce the number of replicas needed to
tolerate $f$ failures from $3f+1$ to $2f+1$. This means an $N$-node network can now tolerate up to $N/2$ adversarial
nodes, as opposed to $N/3$ adversarial nodes in the original PBFT. A2M's and Hybster's safety are dependent on
the trusted code bases (TCBs) that implement simple functions: a log data structure in the former and a
monotonic counter in the latter. 

\subsubsection*{Federated}
Despite numerous improvements to the original protocol, PBFT-based consensus remains communication bound, thus
it ultimately fails to scale beyond a certain number of nodes. To overcome this hard limit without scarifying safety,
Stellar and Ripple adopt an approach that partitions the network into smaller groups called federates. Each federate runs
a local consensus protocol among its members, which does not run into scalability problems because of the small network size. Local
agreements are then propagated to the entire network via nodes lying in the intersections of the federates.
Global consensus can be achieved under certain conditions. For Stellar, the condition is that every two
federates intersect at non-Byzantine nodes. Ripple's safety conditions are that there is a large majority of
honest nodes in every federate, and that the intersection of any two federates contain at least one honest node. 

Both Stellar and Ripple assume federates are pre-defined and their safety conditions can be enforced by a network
administrator. In a decentralized environment where node identities are unknown, such assumptions do not hold.
Byzcoin~\cite{byzcoin} and Elastico~\cite{elastico} propose novel, two-phase protocols that combine PoW and
PBFT. In the first phase, PoW is used to form a consensus group. Byzcoin implements this by having a sliding
window over the blockchain and selecting the miners of the blocks within the window. Elastico~\cite{elastico}
groups nodes by their identities that change every epoch. In particular, a node identity is its solution to a
cryptographic puzzle. In the second phase, the selected nodes perform PBFT to determine the next block. The
end result is faster block confirmation time at a scale much greater than traditional PBFT (over $1000$
nodes). 

Similar to Byzcoin and Elastico, Dfinity~\cite{dfinity} and Algorand~\cite{algorand} select at each round a
random set of nodes that can propose blocks. Unlike the former, they dispense with PoW and instead use
verifiable random functions (VRFs) to select the consensus group. In Dfinity, the VRF is based on the
threshold signature of the previous block. In Algorand, it is based on a random seed published in the previous
block and the node's secret key.  

\subsubsection*{Non-Byzantine}
The systems described so far in this section tolerate Byzantine failures, rendering them attractive for public
settings and for private settings where the cost of engaging trusted parties (for example, for escrowing
assets) is high. Some blockchains, however, assume trusted parties in order to simplify their
designs. These blockchains have no safety guarantees when any of such parties behaves maliciously. 

Openchain~\cite{openchain} relies on a single trusted party (called validator) that determines the next block. Consequently, it is most
vulnerable to attacks as the validator is the single point of failure. Multichain and Parity have more than one trusted
party which is referred to as authority in their systems. Each authority is given a time slice, via round-robin scheduling,
during which it can append new blocks to the chain. This simple proof-of-authority (PoA) protocol avoids single point of
failure while ensuring balanced workloads among the authorities. HydraChain and BigChainDb also have multiple
authorities, but one authority cannot unilaterally decide the next blocks. Instead, the block is decided via majority
voting. Quorum~\cite{quorum} employs Raft~\cite{raft} as the consensus protocol among its authorities.  Raft implements
crash tolerant state machine replication, which is an important building block of modern distributed database
systems. Using Raft, Quorum is able to make safe progress even when some authority nodes crash. 

Corda's consensus protocol is executed by a set of trusted parties called {\em notaries} which check if a given
transaction has been executed before. By delegating this check to an entity outside of the blockchain, Corda
can justify  using Raft for consensus.  Transactions in Corda are sent to the notaries before being confirmed in the
blockchain.  The notaries then use Raft to ensure that the transactions are replicated among themselves and
remain highly available despite crashes. 

The latest release of Hyperledger (v1.0) outsources the consensus component to Kafka --- another building block often
found in distributed database systems. More specifically, transactions are sent to a centralized Kafka service
which orders them into a stream of events. Every node subscribes to the same Kafka stream and therefore is
notified of new transactions in the same order as they are published. Since there is only one  Kafka service,
the observed transaction sequence is the same at every node.  

\subsubsection*{Others}
IOTA~\cite{iota} uses its own consensus protocol called Tangle in which the blocks form a direct acyclic graph (DAG) as
opposed to a chain. In addition, a block in Tangle consists of only one transaction. When appended, the block
must {\em approve} two other blocks creating links to them in the DAG. The block is confirmed when it is
approved by many other blocks. Targeting IoT environments, Tangle's main goal is efficiency and low-cost
payment. Although its security has not been rigorously analyzed, the low values of transactions
(micropayments) in Tangle could in practice discourage Byzantine behavior. 

Kadena~\cite{kadena} proposes an extension to Raft that handles Byzantine failures. It introduces various techniques on top of Raft,
such as message signatures, client verification and incremental hashing. However, like Tangle, it is unclear
whether the protocol guarantees safety and liveness.

\subsection{Smart Contracts}
Recall that a smart contract system can be characterized by its language expressiveness or by its execution environment.
Except for Openchain, IOTA, Ripple and Stellar, all systems listed in Table~\ref{table:statb} let users customize
transaction logics to suit their applications. In the following, we group them by the contract language
expressiveness.   

\subsubsection*{Scripts}
Bitcoin provides approximately 200 opcodes, but many of them are disabled in the latest implementation. Users
can write stack-based programs with the opcodes. The most popular contracts in Bitcoin are related to
multi-signatures. One example is the escrow contract that requires $2$ out of $3$ signatures before a coin can
be released.  The language can also implement bounty-hunting style contracts, for example, one that releases
the reward coins when the pre-image of a hash value is found. 

BigchainDB~\cite{bigchaindb} adopts a more expressive language called {\em crypto-condition}. Developed as part of the
Interledger Protocol project~\cite{ilp}, crypto-condition allows specifying complex boolean expressions over
many types of signatures. A crypto-condition script contains {\em conditions} and {\em fulfillments} which
are treated as inputs and output of the script. The available conditions include {\em timeout} which enables
time-release contracts. Crypto-condition's encoding is higher level than Bitcoin opcodes, making it easy to
express complex logics. 

\subsubsection*{Turing complete}
Ethereum is among the first blockchains offering Turing-complete smart contracts. Users write their contracts in either
Solidity, Serpent or LLC language, which then get compiled to EVM bytecodes. EVM executes normal crypto-currency
transactions, and it treats smart contract bytecodes as a special transaction. Specifically, each smart contract
is given its own memory to store local states. The memory is exposed as a key-value storage, though Solidity
provides high-level data types such as map, array and composite structures. Resources consumed during
execution of the contract, both in terms of CPU and memory, are tracked by EVM and charged to the transaction
sender's account. EVM also keeps track of intermediate state changes and reverse them if there are
insufficient funds to pay for the execution.  

Hyperledger does not have its own bytecotes. Instead, it runs its language-agnostic smart contracts inside
Docker containers.  Specifically, a contract can be written in any language, which is then compiled into
native code and packed into a Docker image. When the contract is uploaded, each node starts a new container
with that image. Invoking the contract is done via Docker APIs. The contract can access the blockchain states
via two methods {\em getState} and {\em putState} exposed by a shim layer. One benefit of Hyperledger is that
it supports multiple high-level programming languages like Go and Java. However, its key-value interfaces with
the blockchain necessitates extra application logics for mapping high-level data structures into key-value tuples.  

Sawtooth Lake supports smart contracts in the form of transaction families. Each family is a user-defined Python class
loaded into the ledger during start up. The contract is executed in the native runtime environment as a normal Python program. 

One consequence of supporting Turing complete contracts is that software bugs are all but inevitable. While
empowering, the Ethereum smart contract model receives strong criticism because it directly exposes Ethers
against programming bugs. The security concerns indeed materialized in the DAO attack~\cite{dao} in which the
attacker stole \$$50M$ worth of asset. The attack exploits a concurrency bug in the DAO smart contract which
allows one to repeatedly draw more money than what is specified in the transaction. Such bugs are inherent in
a language like EVM which has weak or no formal specifications of its semantics. OYENTE~\cite{smarter-contract}
presents three major causes of security bugs: transaction order dependencies, timestamp dependencies and
mishandled exceptions.  It formalizes Ethereum semantics and proposes a tool for checking bugs directly on
EVM bytecodes. The tool discovered over $8000$ Ethereum contracts (worth over \$$60M$) with potential
security bugs. 

Like any other transactions on the blockchain, smart contract executions are transparent. It means the inputs,
outputs and the states of the contract are visible to the network. Hawk~\cite{kosba16} extends Zerocash to
provide transaction privacy for smart contracts.  The main challenge compared to Zerocash lies in the
arbitrary transaction logics, whereas in Zerocash the logics are constrained by a small set of operations.
Another challenge is to protect local states, which is not applicable in Zerocash. Given a contract, Hawk
compiles it with zkSNARK to make it privacy preserving. Transaction inputs and outputs are pre- and
post-processed via Hawk to hide the complex cryptographic details. Although the protocols incur large overhead
both in time and space, Hawk represents a practical cryptographic system that achieves both transaction privacy and
fairness.   

\subsubsection*{Verifiable}
Even before the DAO attack, some blockchains have rejected the models that allow for unconstrained
computations. The languages of Kadena, Tezos and Corda are more powerful than Bitcoin scripts, but they trade
Turing completeness for safety. Kadena's language is a Lisp-like functional language called
Pact~\cite{kadena}. A Pact contract is stored in the ledger in human readable form, which is then parsed and
executed in Ocaml. It is strongly typed and can be formally verified.  Similarly, Tezos's stack-based language
called Michelson comes with a strong type system and fully specified semantics.  As a result, Tezos contracts
can be statically checked for safety. In Corda, a contract is a sequence of {\em pure functions} that do not
modify the states.  Because the functions are merely constraints, the contract's safety can be formalized and
verified. 

%% file: bb.tex
\section{BLOCKBENCH}
\label{sec:blockbench}
The previous section has presented a thorough qualitative analysis of existing blockchains. In this section,
we describe our benchmarking framework called \name~\cite{blockbench-sigmod}.  Designed for quantitative
analysis of blockchains as data processing platforms, the framework targets private blockchains with
Turing-complete smart contracts. \name\ is open source~\cite{blockbench} and contains data processing
workloads commonly found in database benchmarks.  

\subsection{Layers}
\label{subsec:layers}
\begin{figure}
\centering
\includegraphics[width=0.53\textwidth]{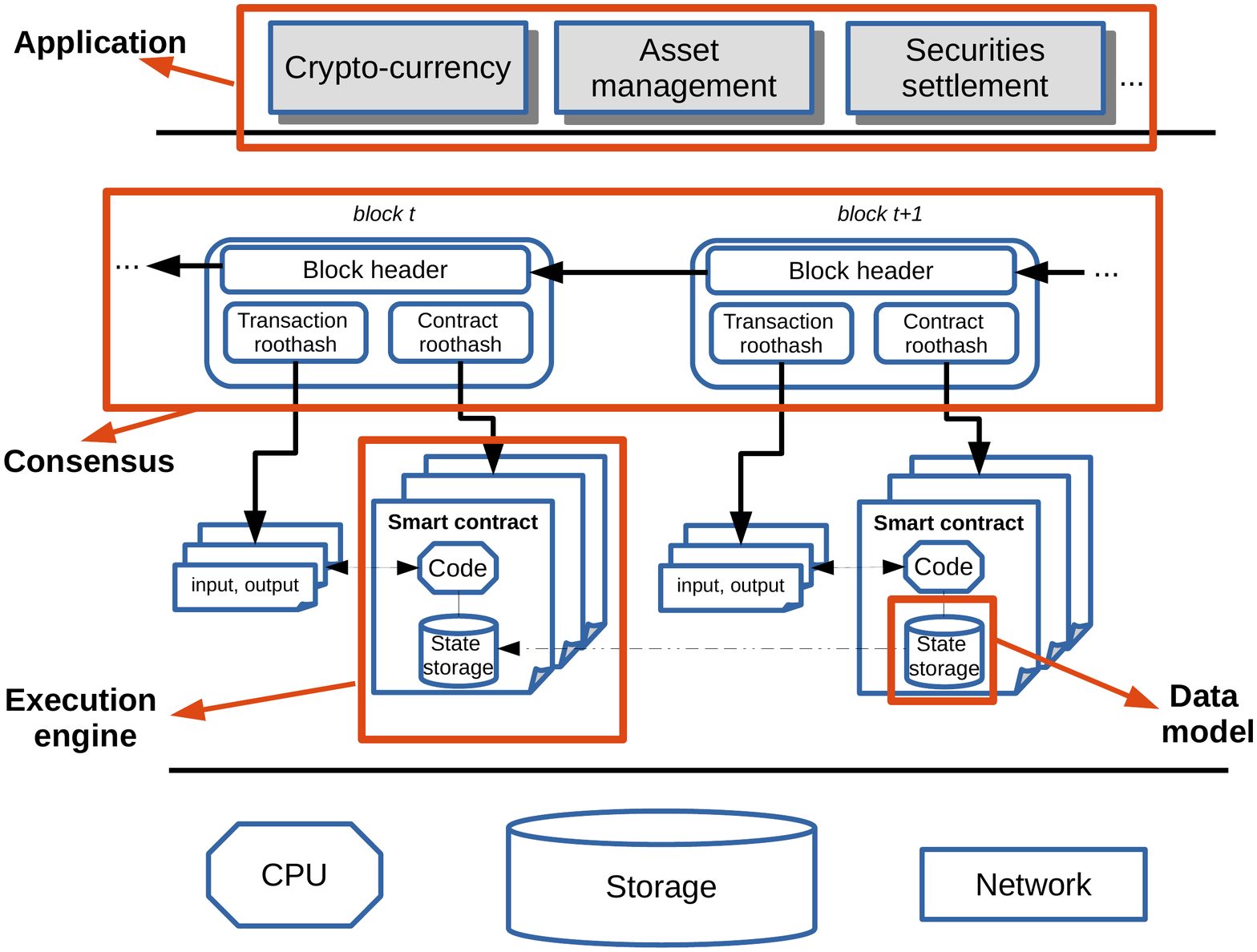}
\caption{Blockchain software stack on a fully validating node.}
\label{fig:stack_annotated}
\end{figure}

\begin{figure}
\centering
\includegraphics[width=0.48\textwidth]{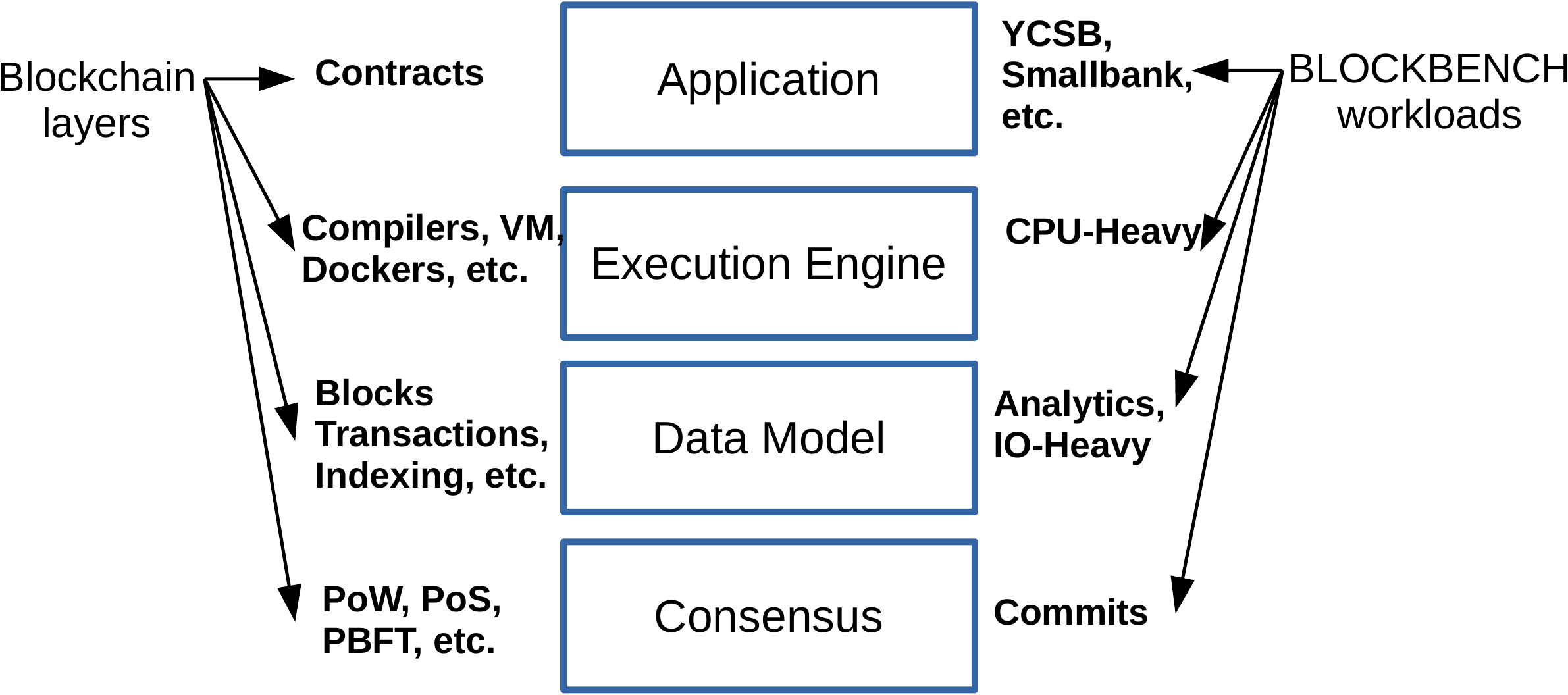}
\caption{Blockchain software layers and corresponding benchmark workloads.}
\label{fig:layers}
\end{figure}
\name\ targets blockchains that function as data processing platforms. Such a blockchain must have
no restrictions on the application logics, thus it must support Turing complete smart contracts.
Figure~\ref{fig:stack_annotated} shows the logical components of the blockchain software stack, from which we
refine the taxonomy described in Section~\ref{sec:tax} into four concrete layers shown in
Figure~\ref{fig:layers}. For each layer there are multiple \name\ workloads for evaluating it individually. 

The consensus layer implements the consensus protocol. The data model layer contains the structure, content
and operations on the blockchain data. The execution layer includes details of the runtime environment for
executing smart contracts. Finally, the application layer includes classes of blockchain applications. Croman
et. al. ~\cite{croman2016scaling} proposed to divide blockchain into several planes: network, consensus,
storage, view and side plane. While similar to \name's four layers, the plane abstraction was geared towards
crypto-currency applications and did not take into account the execution of smart contracts. 

\subsection{Implementation}

\name\ stack consists of a frontend interface for integrating new benchmark workloads, a backend interface for
integrating new blockchains, and a driver for driving the workloads. A new blockchain can be integrated into the
framework's backend by implementing the {\tt IBlockchainConnector} interface. The interface contains operations for
deploying the smart contract application, invoking it by sending a transaction, and for querying the blockchain states.
Ethereum, Parity and Hyperledger are the current backends, while ErisDB (or Monax), Quorum and Sawtooth Lake integration
are under development. A new benchmarking workload can be added by implementing {\tt IWorkloadConnector}
interface\footnote{We assume that the smart contract implementing the workload's logic is already implemented and
deployed on the blockchain.}. The {\tt Driver} takes as input a workload and sends transactions to the blockchain
according to user-defined configurations (number of operations, number of clients, threads, etc.).  It collects runtime
statistics which are used to compute five important metrics. 
\begin{itemize}
\item Throughput: the number of successful transactions per second. A workload can be configured
with multiple clients and threads per clients to saturate the blockchain throughput. 
\item Latency: the response time per transaction. {\tt Driver} implements blocking transactions, i.e. it
waits for one transaction to finish before starting another. 
\item Scalability: the  changes in throughput and latency when increasing the number of nodes and number of
concurrent workloads.  
\item Fault tolerance: the changes in throughput and latency during node failure. We simulate crashes, network
delays and random message corruptions. 
\item Security metrics: the ratio between the total number of blocks included in the
main branch and the total number of confirmed blocks. The lower the ratio, the less vulnerable the system
is from double spending or selfish mining.  
\end{itemize}


\subsection{Workloads}
\label{subsec:workloads}
\name\ comes with macro benchmark workloads for evaluating the application layer, and micro benchmark workloads for
analyzing the lower layers. Smart contract implementations of the workloads shown in
Figure~\ref{fig:layers} are available and can be readily deployed on Ethereum, Parity and Hyperledger. 

\subsubsection*{Macro benchmark workloads} 
We port two popular database benchmark workloads into \name, namely YCSB and Smallbank. YCSB is widely used for
evaluating NoSQL databases, for which we implement a simple smart contract which functions as a key-value storage.
The {\tt WorkloadClient} is based on the YCSB driver~\cite{ycsb} which preloads each storage with a number of
records, and supports requests with different ratios of read and write operations. For Smallbank~\cite{smallbank},  a
popular benchmark for OLTP workload, we implement a smart contract that transfers money from one account to
another.  

Besides database workloads, \name\ also provides three other workloads based on real Ethereum contracts. The first is {\em
EtherId}, a popular contract implementing a domain name registrar. The second is {\em Doubler}, the pyramid scheme
contract shown earlier in Figure~\ref{fig:doubler}. The third is {\em WavesPresale} that implements a crowdfunding campaign via
digital token sales. 

\subsubsection*{Micro benchmark workloads}
For the consensus layer, \name\ provides {\em DoNothing} workload in which the smart contract accepts a transaction as
input and simply returns. Since the contract execution involves minimal number of operations at the execution and data
model layer, the overall performance will be determined by the consensus layer.  

\begin{figure}
\centering
{\footnotesize
\begin{verbatim}
type account_t struct {
    Balance     int
    CommitBlock int
}
type transaction_t {
    From string
    To   string
    Val  int
}
func Invoke_SendValue(from_account string,
        to_account string, value int) {
    var pending_list []transaction_t
    pending_list = decode(GetState("pending_list"))
    var new_txn transaction_t
    new_txn = transaction_t {
        from_account, to_account, value
    }
    pending_list = append(pending_list, new_txn)
    PutState('pending_list', encode(pending_list))
}
func Query_BlockTransactionList(block_number int)
     []transaction_t {
    return decode(GetState("block:"+block_number))
}
func Query_AccountBlockRange(account string,
        start_block int, end_block int)
     []account_t {
    version := decode(GetState(account+":latest"))
    var ret []account_t
    while true {
        var acc account_t
        acc = decode(GetState(account+":"+version))
        if acc.CommitBlock >= start_block &&
           acc.CommitBlock < end_block {
            ret = append(ret, acc)
        } else if acc.CommitBlock < start_block {
            break;
        }
        version -= 1
    }
    return ret
}
\end{verbatim}
}
\caption{Code snippet from the VersionKVStore smart contract for Analytics workload (Q1 and Q2).}
\label{fig:analytic_sc} 
\end{figure}

For the data model layer, \name\ provides {\em Analytics} workload that is similar to an OLAP workload. In
particular, it performs scan-like and aggregate queries whose performance is determined by the system's data
model. Specifically, there are two queries: 

\begin{itemize} \item [Q1:] {\em Compute the total transaction values committed between block i and block j}. 
\item [Q2:] {\em Compute the largest transaction value involving a given state (account) between block i and block
j}.  
\end{itemize}
For Ethereum and Parity, both queries can be implemented via JSON-RPC APIs that return transaction details and account
balances at a specific block. For Hyperledger, however, the second query must be implemented via a smart
contract (VersionKVStore), because Hyperledger has no APIs for querying historical states. Figure
~\ref{fig:analytic_sc} shows the contract implementation in Hyperledger. To support historical data lookup,
the contract appends a counter to the key of each account. To fetch a specific version of an account, the key
{\tt account:version} is used. The latest version is stored at the key {\tt account:latest}. The contract also
keeps keep a {\tt CommitBlock} value in the data field for every version to point to the block number in which
the current version is committed. To fetch the balances of a given account in a given block range, the
contract scans all versions of this account and returns the corresponding balance when the version's
CommitBlock value is in the specified range. 

Another workload for the data model layer stresses the persistent storage. In particular, the {\em IOHeavy} workload
evaluates the blockchain's IO performance by invoking a contract that performs a large number of random writes and random
reads to the local states. 

Finally, for the execution layer \name provides the {\em CPUHeavy} workload. It measures the efficiency of the
execution layer for computationally heavy tasks by invoking a contract that executes quick sort algorithm over
a large array. 

%% file: benchmark.tex
\section{Evaluation}
\label{sec:evaluation}
We selected Ethereum, Parity and Hyperledger for a comparative study using \name. They occupy different
positions in the design space, and are considered the most mature in terms of the codebase
and user base. We used the popular Go implementation of Ethereum, {\em geth v1.4.18}, the Parity release {\em
v1.6.0}. Unless otherwise specified, the Hyperledger version is {\em v0.6.0-preview}. We set up a private
testnet for Ethereum and Parity by defining a genesis block and directly adding peers to the miner network.
For Ethereum, we manually tuned the {\tt difficulty} variable in the genesis block to ensure that miners do
not diverge in large networks. For Parity, we set the {\tt stepDuration} variable to 1. In both Ethereum and
Parity, {\tt confirmationLength} is set to $5$ seconds. The default batch size in Hyperledger is $500$. 

The experiments were run on a 48-node commodity cluster. Each node has an E5-1650 3.5GHz CPU, 32GB RAM, 2TB hard drive,
running Ubuntu 14.04 Trusty, and connected to the other nodes via 1GB switch. For Ethereum, we reserved 8
cores out of the available 12 cores per machine, so that the periodic polls from the client's driver process do
not interfere with the mining process. Our main findings are as follows:

\begin{itemize} 
\item Hyperledger performs consistently better than Ethereum and Parity across the benchmarks. But it fails to scale up
to more than $16$ nodes. 
\item Ethereum and Parity are more resilient to node failures, but they are vulnerable to security attacks that forks
the blockchain. 
\item The main bottlenecks in Hyperledger and Ethereum are the consensus protocols, but for Parity the bottleneck is
caused by transaction signing. 
\item Ethereum and Parity incur large overheads in terms of memory and disk usage. Their execution engine is also less
efficient than that of Hyperledger. 
\item Hyperledger's data model is low level, but its flexibility enables customized optimization for analytical queries.   
\end{itemize}

\subsection{Macro benchmarks}
This section discusses the performance of the blockchains at the application layer, using YCSB and Smallbank
benchmarks. 

\subsubsection*{Throughput and latency}
\begin{figure}
\includegraphics[width=0.45\textwidth]{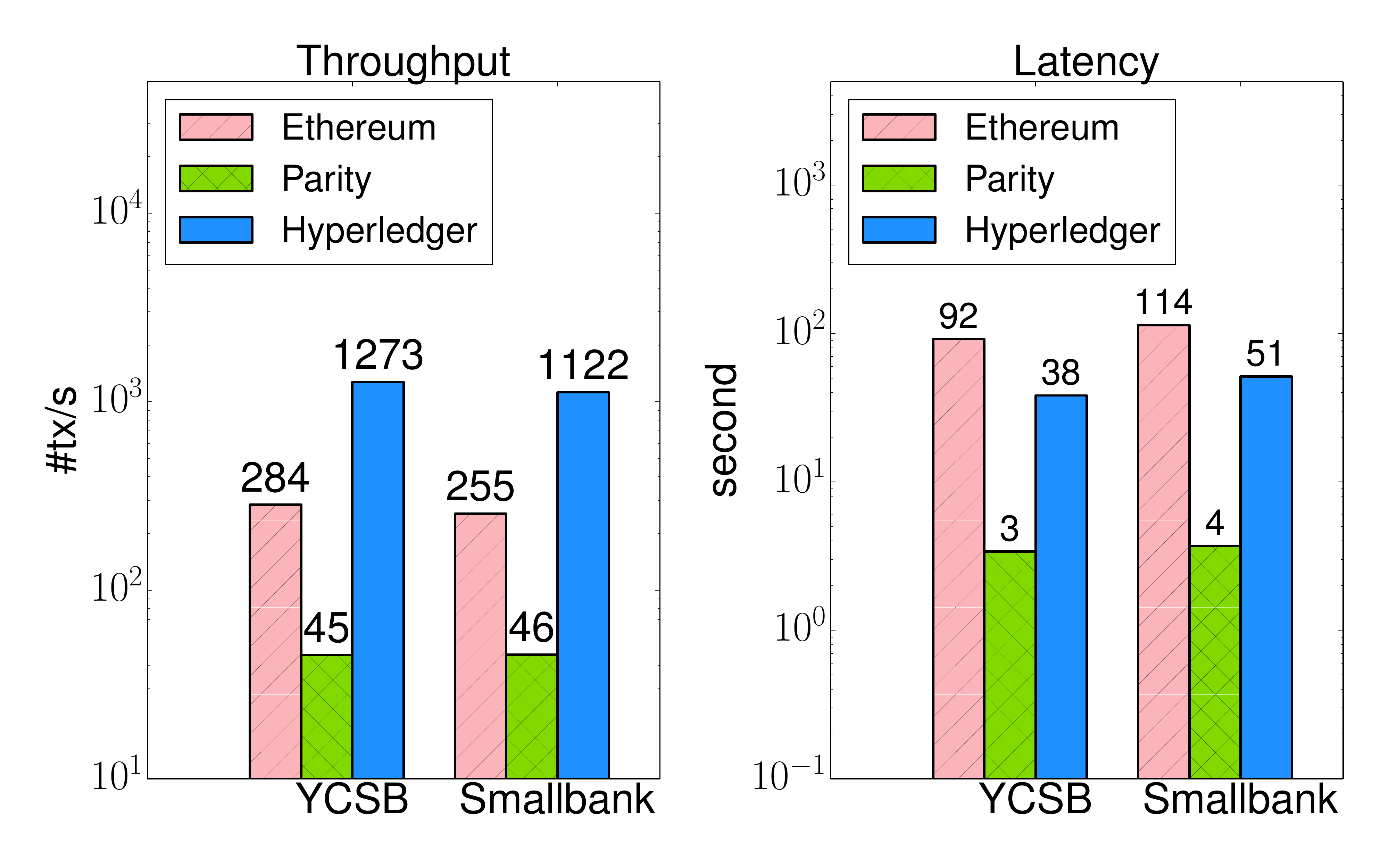}
\caption{Blockchain peak performance with 8 clients and 8 servers. The benchmark workload is YCSB. The performance
against Smallbank workload is similar.}
\label{fig:saturation}
\end{figure}

Figure~\ref{fig:saturation} shows the peak performance with 8 servers and 8 concurrent clients over the
period of 5 minutes. We observe that in terms of throughput, Hyperledger outperforms the other two in both
benchmarks. The gap between Hyperledger and Ethereum is due to the difference in the consensus protocols: one
is based on PBFT while the other is based on PoW. With 8 servers, the communication cost from broadcasting
messages is cheaper than block mining whose difficulty is set at roughly $2.5s$ per block. The gap between
Parity and Hyperledger is not due to consensus protocols, as Parity's PoA protocol is expected to be simpler
and more efficient than both PoW and PBFT.  Instead, we observe that Parity processes transactions at a
constant rate, and that it enforces a maximum client request rate at around $80$ tx/s.

\begin{figure}
\centering
\includegraphics[width=0.45\textwidth]{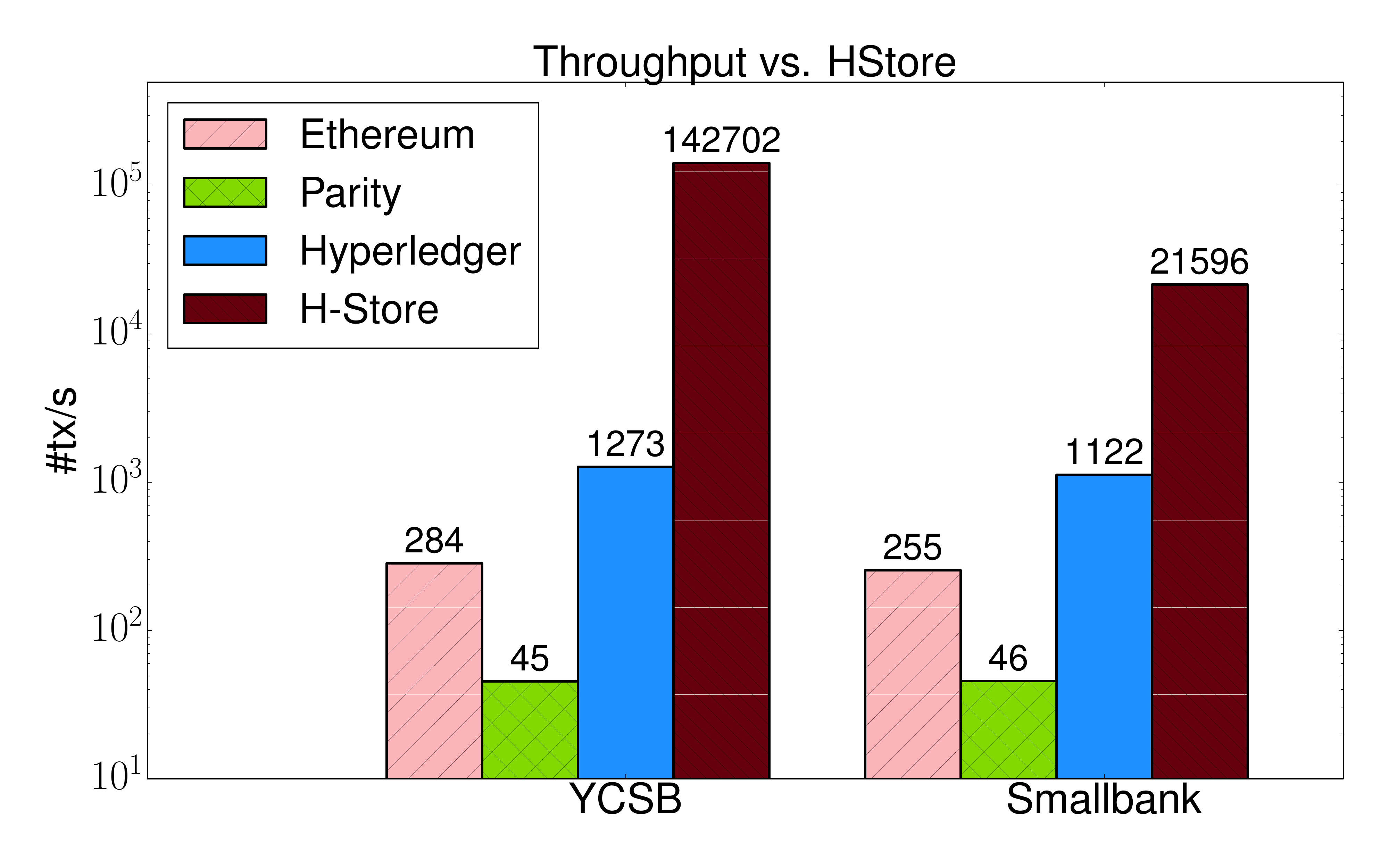}
\caption{Performance of the three blockchain systems versus H-Store.}
\label{fig:vs_hstore}
\end{figure}

To put their performance in context, we compare the three blockchains against a popular in-memory database system,
namely H-Store, using the YCSB and Smallbank workload. Blockchains and databases do not
necessarily share the same design goal: the former are not designed for general data processing, nor do the
latter protect data integrity against Byzantine failures. Nonetheless, we argue that the comparison offers useful
insights into the design trade-offs and relative performance of the two systems. We ran H-Store's own
benchmark driver and set the transaction rate at 100,000 tx/s. Figure~\ref{fig:vs_hstore} shows at least an
order of magnitude gap in throughput and two order of magnitude in latency. Specifically, H-Store achieves
over 140K tx/s throughput while maintaining sub-millisecond latency.  The gap in performance is due to the
cost of consensus protocols.  For YCSB, for example, H-Store requires almost no coordination among peers,
whereas Ethereum and Hyperledger suffer the overhead of PoW and PBFT.  An interesting observation is the
overhead of Smallbank. Recall that compared to YCSB, Smallbank consists of more complex transactions in which
multiple keys are updated in a single transaction. Smallbank is simple but is representative of the large
class of transactional workloads such as TPC-C. We observe that in H-Store, Smallbank achieves $6.6$x lower
throughput and $4$x higher latency than YCSB, which reflects the cost of distributed transaction management.
In contrast, the blockchains suffer modest degradation in performance: $10\%$ in throughput and $20\%$ in
latency. This is because each node in the blockchains maintains the complete states, therefore it pays no
overhead in coordinating distributed transactions since the states are not partitioned.    

\subsubsection*{Scalability}
\begin{figure}
\centering
\includegraphics[width=0.45\textwidth]{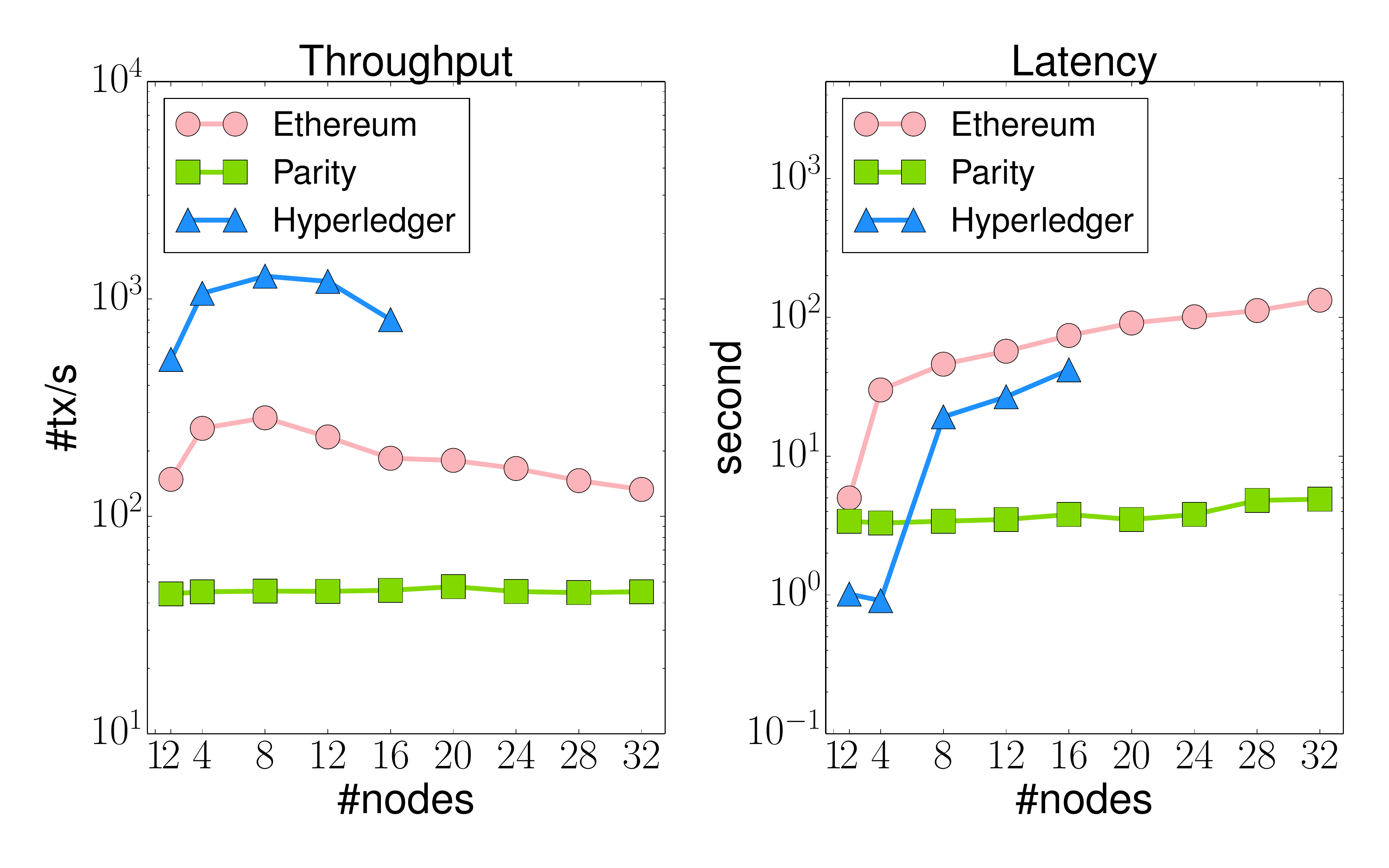}
\caption{Performance scalability (with the same number of clients and servers). The benchmark workload is
YCSB. The scalability against Smallbank is similar, except that Hyperledger fails beyond 8 nodes instead of
16.}
\label{fig:scale}
\end{figure}

\begin{figure}
\centering
\includegraphics[width=0.45\textwidth]{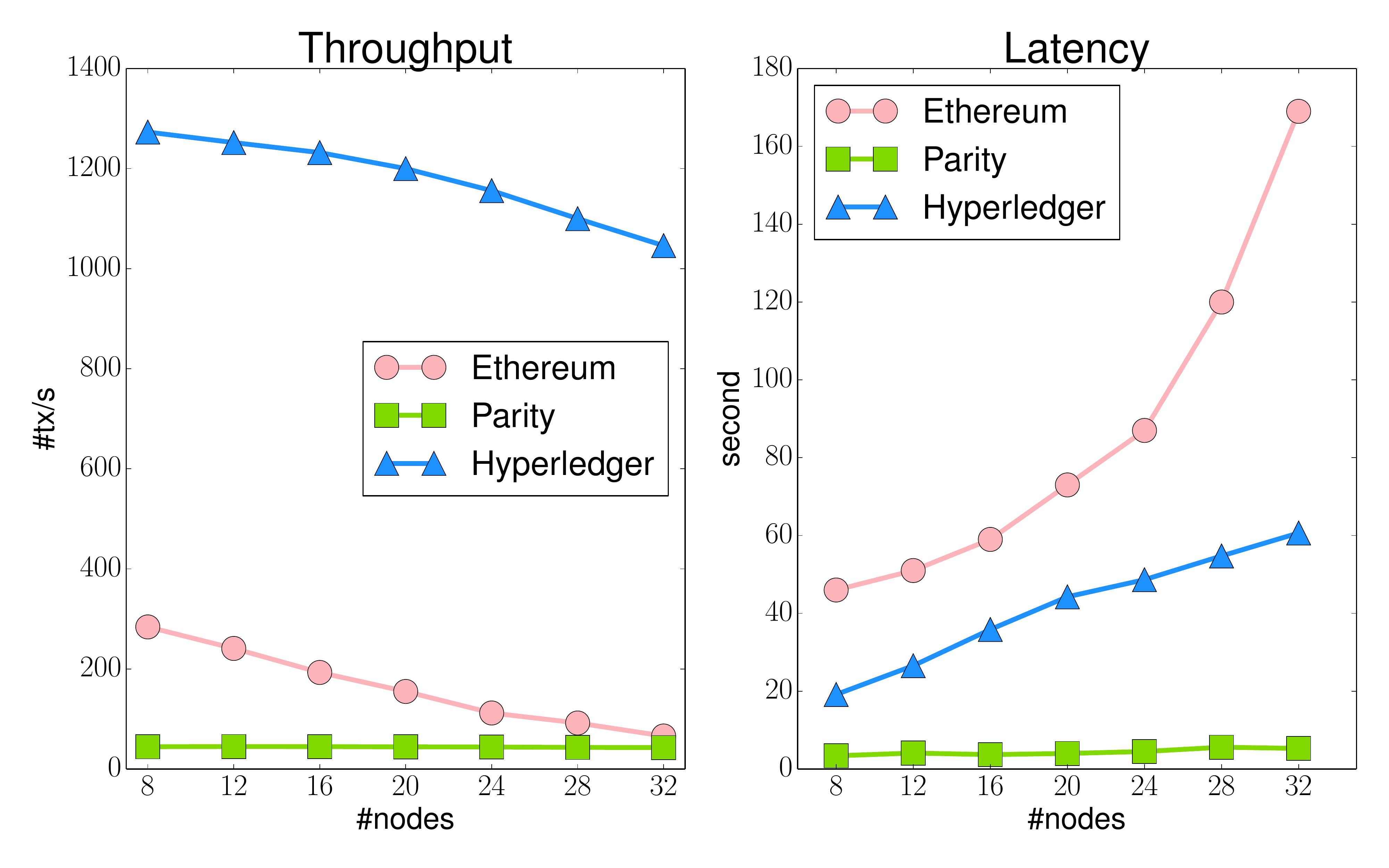}
\caption{Performance scalability (with 8 clients).}
\label{fig:scale_fixed}
\end{figure}

We fixed the client request rate ($320$ requests per second for Hyperledger, $160$ requests per second for
Ethereum and Parity) and increased both the number of clients and the number of servers.
Figure~\ref{fig:scale} illustrates how well the three systems scale to handle larger YCSB workloads.  Parity's
performance remains constant as the network size and offered load increase, due to the constant transaction
processing rate at the servers.  Interestingly, while Ethereum's throughput and latency degrade almost
linearly beyond 8 servers, Hyperledger stops working beyond 16 servers.

To understand why Hyperledger failed to scale beyond 16 servers and 16 clients, we examined the system logs
and find that the nodes never reached consensus on any batch of transactions. We observe a large number of
messages being dropped even when there are fewer than 16 servers and clients. Furthermore, the servers repeatedly
triggered view changes but never succeeded. At the client side, requests took longer to return as time passed,
suggesting that the server or the network were saturated. Since the original PBFT protocol guarantees both
liveness and safety, we can attribute this failure to scale to Hyperledger's implementation. Further
investigation reveals that it is indeed the case.

\vspace{0.5cm}
Hyperledger uses gRPC for communication between servers. Each server keeps a separate message queue for every
other servers in the network. The queue size is defined during initial setup (the default size being 1000
messages) and the default behavior is to drop messages when the queue is full. In the current design, both
client requests (transactions) and consensus messages (pre-prepare, prepare, commit, view changes) are sent on
the same channel, that is they share the same queue. For large numbers of concurrent clients and servers, the
channels are dominated by client requests, increasing the probability of consensus messages being dropped.
Without a sufficient number of consensus messages, either a batch timer or a view-change timer will expire. In
the first case, the PBFT leader resends messages of the current consensus round. In the second, the servers
start the view change phase which broadcasts multiple rounds of consensus messages. As client requests are
still occupying the network channels, both consensus or view change messages are dropped with high probability.
Consequently, the network gets stuck in perpetual attempts to establish a stable view. The fact that PBFT is
sensitive to network conditions has been observed in the past~\cite{clement09}.

We note that in its latest release ($v1.0$) Hyperledger has replaced PBFT with a global ordering service.
Implemented using Kafka, this new consensus engine may offer higher throughput than PBFT, but it offers no
protection against Byzantine failures.  


The results so far indicate that scaling both the number of clients and number of servers degrades the performance
and even causes Hyperledger to fail. We next examined the costs of increasing the number of servers alone while fixing
the number of clients to 8. Figure~\ref{fig:scale_fixed} shows that the performance becomes worse as there are more servers,
meaning that the systems incur some network overheads. For Hyperledger, having more servers means more
messages being exchanged and higher overheads. In particular, to a confirm a batch of transaction in a larger
network, the leader in Hyperledger needs to wait for larger sets of messages, therefore lowering overall
throughputs. We note that with a fixed number of clients Hyperledger can scale up to 32 nodes, as oppose to
failing after 16 nodes as in Figure~\ref{fig:scale}. This is because with fewer clients, the message queues at
each node are not saturated with client requests and therefore consensus messages are less likely to get
dropped. 

For Ethereum, even though it is computation bound, it still consumes a modest amount of network resources for
propagating transactions and blocks to other nodes. Furthermore, with larger network, the difficulty is
increased to account for the longer propagation delays. We observe that to prevent
the network from diverging, the difficulty level increases at a higher rate than the number of nodes. Thus, one
reason for Ethereum's throughput degradation is due to network sizes.  Another reason is that in our settings,
$8$ clients send requests to only $8$ servers, but these servers do not always broadcast transactions to each
other (they keep mining on their own transaction pool). As a result, the network mining capability is not
fully utilized. 

\subsubsection{Fault tolerance and security}
\begin{figure}
\centering
\includegraphics[width=0.45\textwidth]{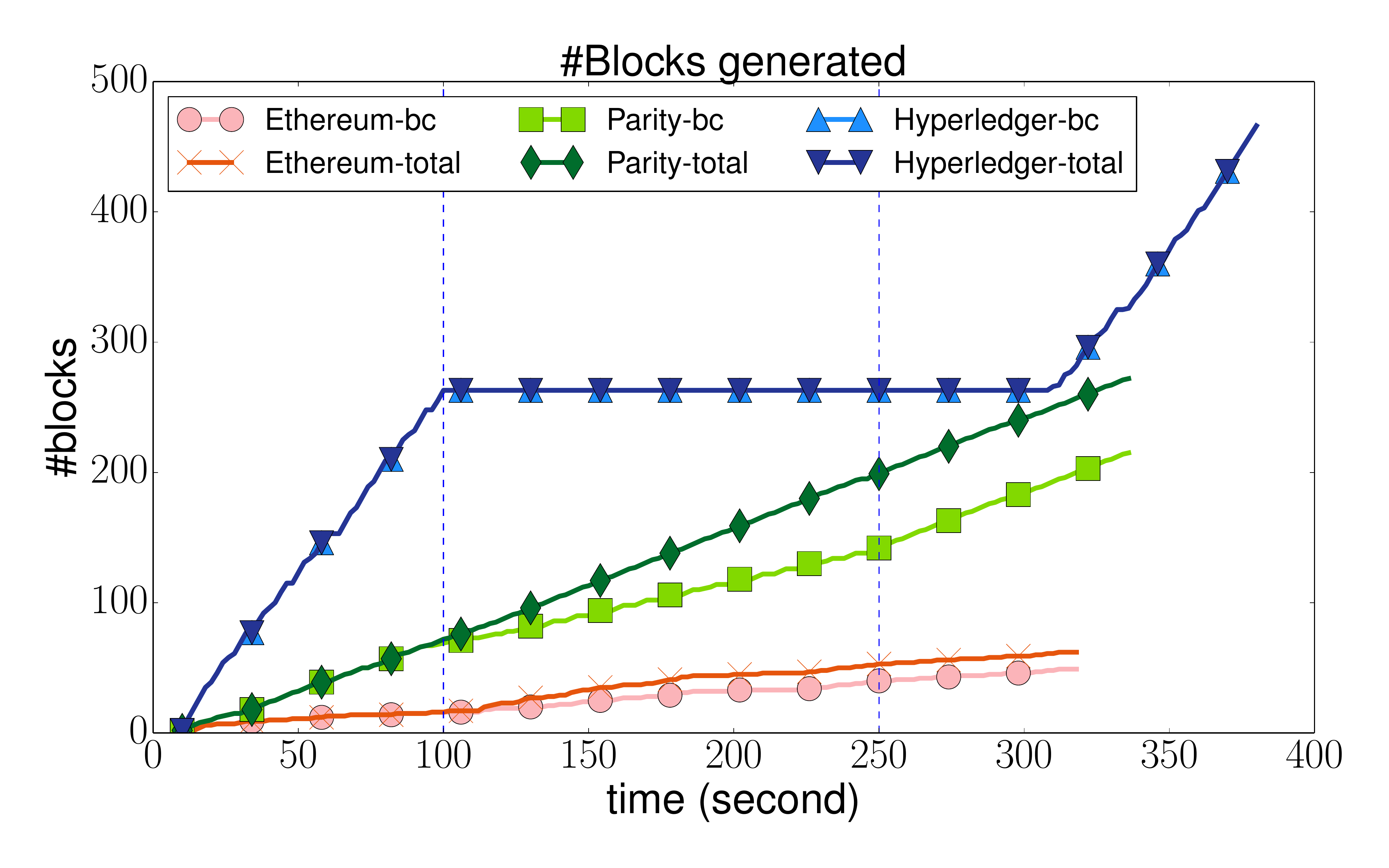}
\caption{Blockchain forks caused by attacks that partitions the network in half at $100^{\text{th}}$ second and lasts for
$150$ seconds. {\em X-total} means the total number of blocks generated in blockchain {\em X},
{\em X-bc} means the total number of blocks that reach consensus in blockchain {\em X}.} \label{fig:security}
\end{figure}

To evaluate how resilient the systems are to crash failures, we ran the systems with 12 and 16 servers, with 8
clients for over 5 minutes, during which we killed off 4 servers at $250^{th}$ second. Due to space
constraints, we only highlight key findings here and refer readers to~\cite{blockbench-sigmod} for more
details.  First, Ethereum is unaffected by the change, suggesting that the failed servers do not
contribute significantly to the mining process. Second, Parity's throughput is also unaffected. It is because
each node is given equal time slice during which it can generate block, thus failing $4$ nodes in Parity
means that the remaining 8 nodes are given bigger time slices. Third , Hyperledger stops generating blocks
after the failure in the 12-server network, which is as expected because the PBFT can only tolerate fewer than
4 failures in a 12-server network.  In the 16-server network, Hyperledger still generates blocks but at a
lower rate, which were caused by the remaining servers having to stabilize the network after the failures by
synchronizing their views.  

We next simulated the attack that could make the blockchains vulnerable to double spending. The attack
essentially creates network partition at $100^{th}$ second that lasts for $150$ seconds. We set the partition
size to be half of the original\footnote{We note that partitioning a $N$-node network in half does not mean
there are $N/2$ Byzantine nodes. In fact, Byzantine tolerance protocols do not count network adversary as
Byzantine failure}. Figure~\ref{fig:security} compares the vulnerability of the three blockchains with $8$
clients and $8$ servers. Recall that vulnerability is measured as the differences in the number of total
blocks and the number of blocks on the main branch. We refer to this as $\Delta$. Both Ethereum and Parity
fork at $100^{th}$ seconds,
and $\Delta$ increases as time passes. For the attack duration, upto $30\%$ of the blocks are generated in the
forked branch, meaning that the systems are highly exposed to double spending or selfish mining attacks. When
the partition heals, the nodes come to consensus on the main branch and discard the forked blocks.  As a
consequence, $\Delta$ stops increasing shortly after $250^{th}$ second. Hyperledger, in stark contrast, has no
fork which is as expected because its consensus protocol is proven to guaranteed safety. We note, however,
that Hyperledger takes longer than the other two systems to recover from the attacks (about $50$ seconds
more). This is because of the synchronization protocol performed after the partitioned nodes reconnect.  In
particular, when the nodes reconnect they enter the view change phase and exchange checkpointed states with
each other in order to establish a new, stable view.  

\subsection{Micro benchmarks}
This section discusses the performance of the blockchains at the execution, data model and consensus layer.  For the
first two layers, the workloads were run with one client and one server. For the consensus layer, 8
clients and 8 servers were used.  


\begin{figure}
\centering
\includegraphics[width=0.45\textwidth]{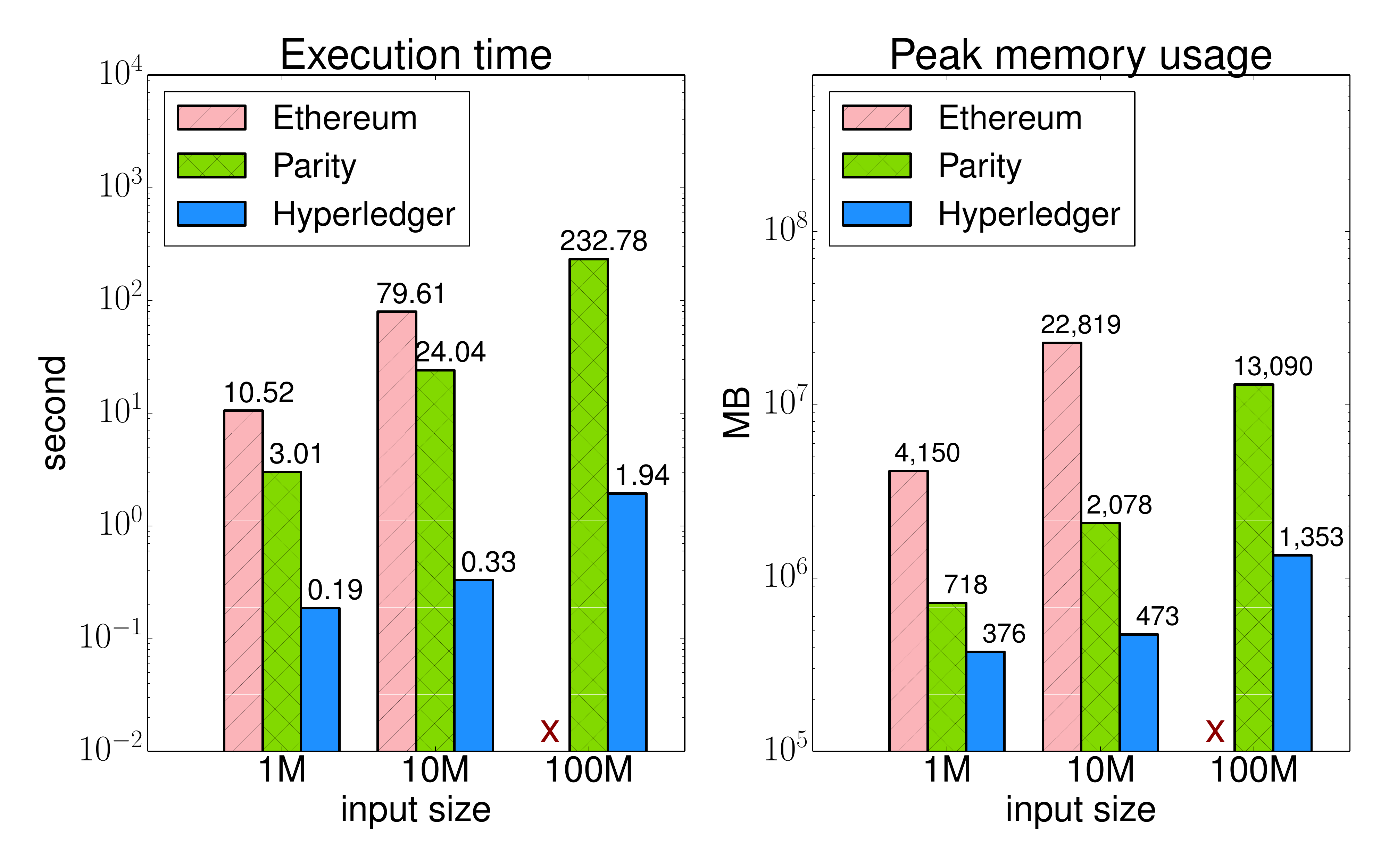}
\caption{CPUHeavy workload, {\bf `X'} indicates Out-of-Memory error.}
\label{fig:cpuheavy}
\end{figure}

\subsubsection*{Execution layer}
We deployed the CPUHeavy smart contract that is initialized with an integer array of a given size. The array is
initialized in descending order. We invoked the contract to sort the array using quicksort algorithm, and measured the
execution time and server's peak memory usage. The results for varying input sizes are shown in
Figure~\ref{fig:cpuheavy}.  Although Ethereum and Parity use the same execution engine, i.e. EVM, Parity's
implementation is more optimized, therefore it is more computation and memory efficient.  An interesting finding is that
Ethereum incurs large memory overhead. In sorting $10M$ elements, it uses $22$GB of memory, as compared to $473$MB used
by Hyperledger. Ethereum runs out of memory when sorting more than $10$M elements.  In Hyperledger, the smart contract
is compiled and runs directly on the native machine within Docker environment, thus it does not have the overheads
associated with executing high-level EVM byte code. As the result, Hyperledger is much more efficient in term of speed
and memory usage.  Finally, we note that all three systems fail to make use of the multi-core architecture, i.e. they
execute the contracts using only one core.

\begin{figure*}
\centering
\subfloat[Write]{\includegraphics[width=0.33\textwidth]{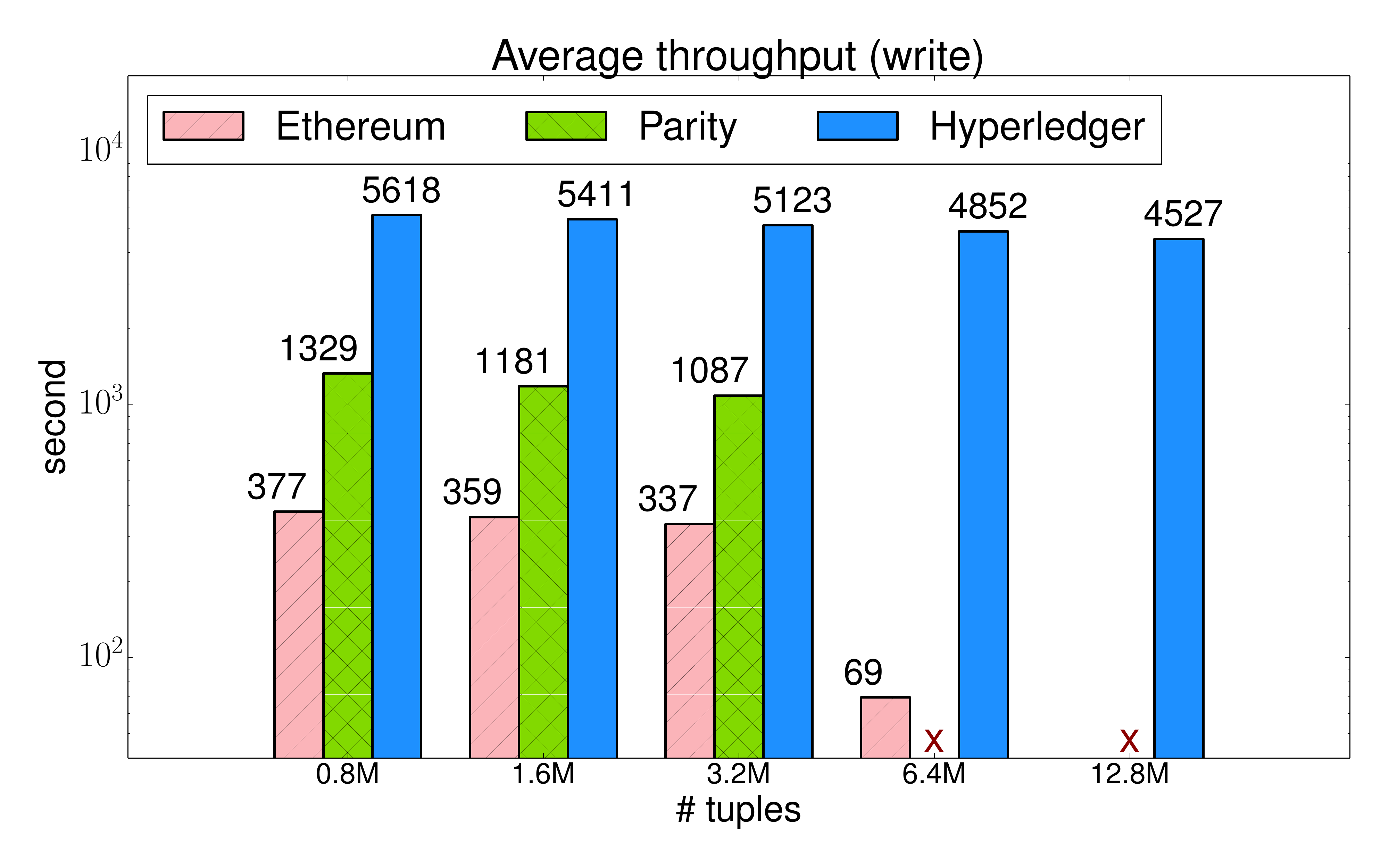}}
\subfloat[Read]{\includegraphics[width=0.33\textwidth]{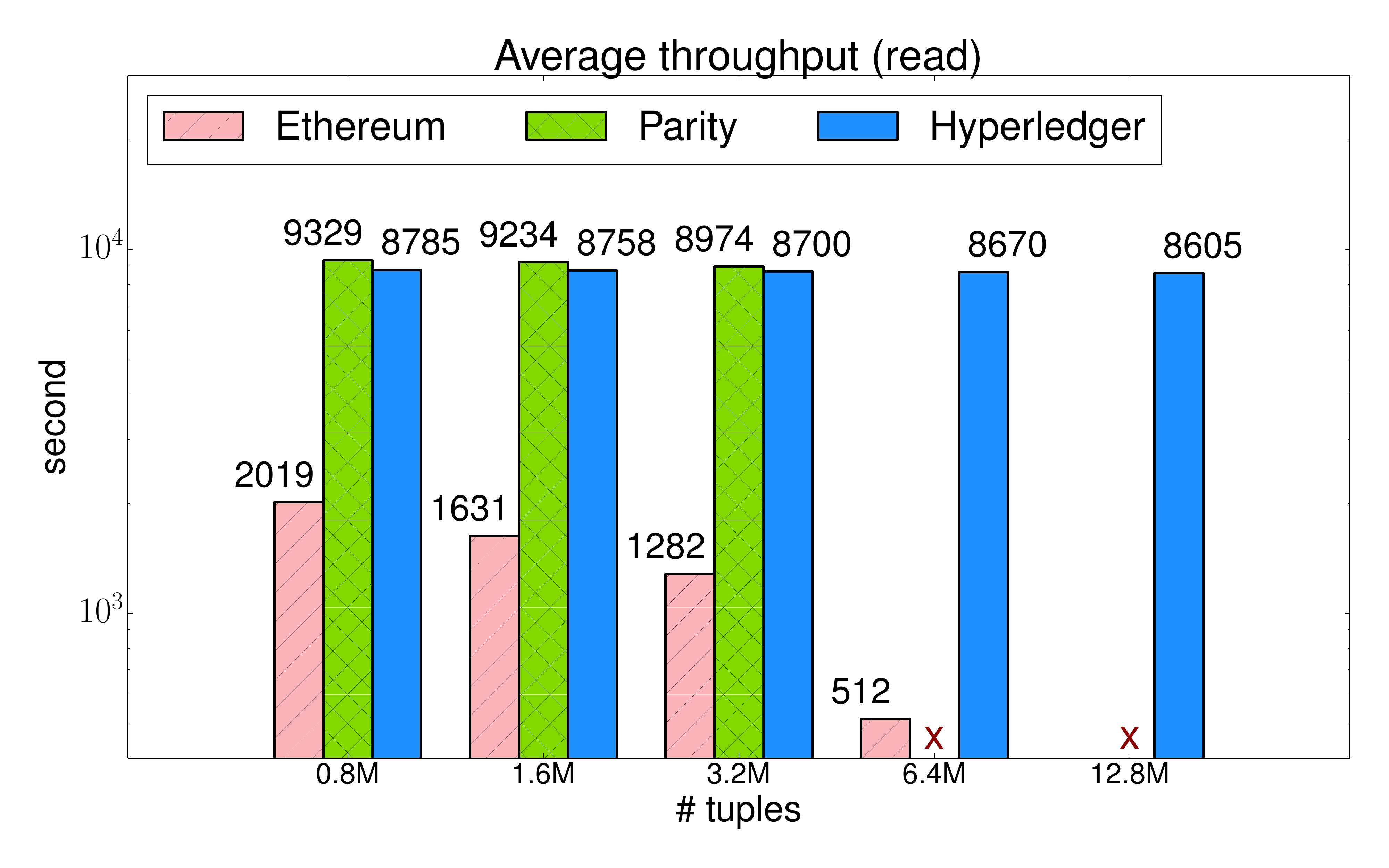}}
\subfloat[Disk usage]{\includegraphics[width=0.33\textwidth]{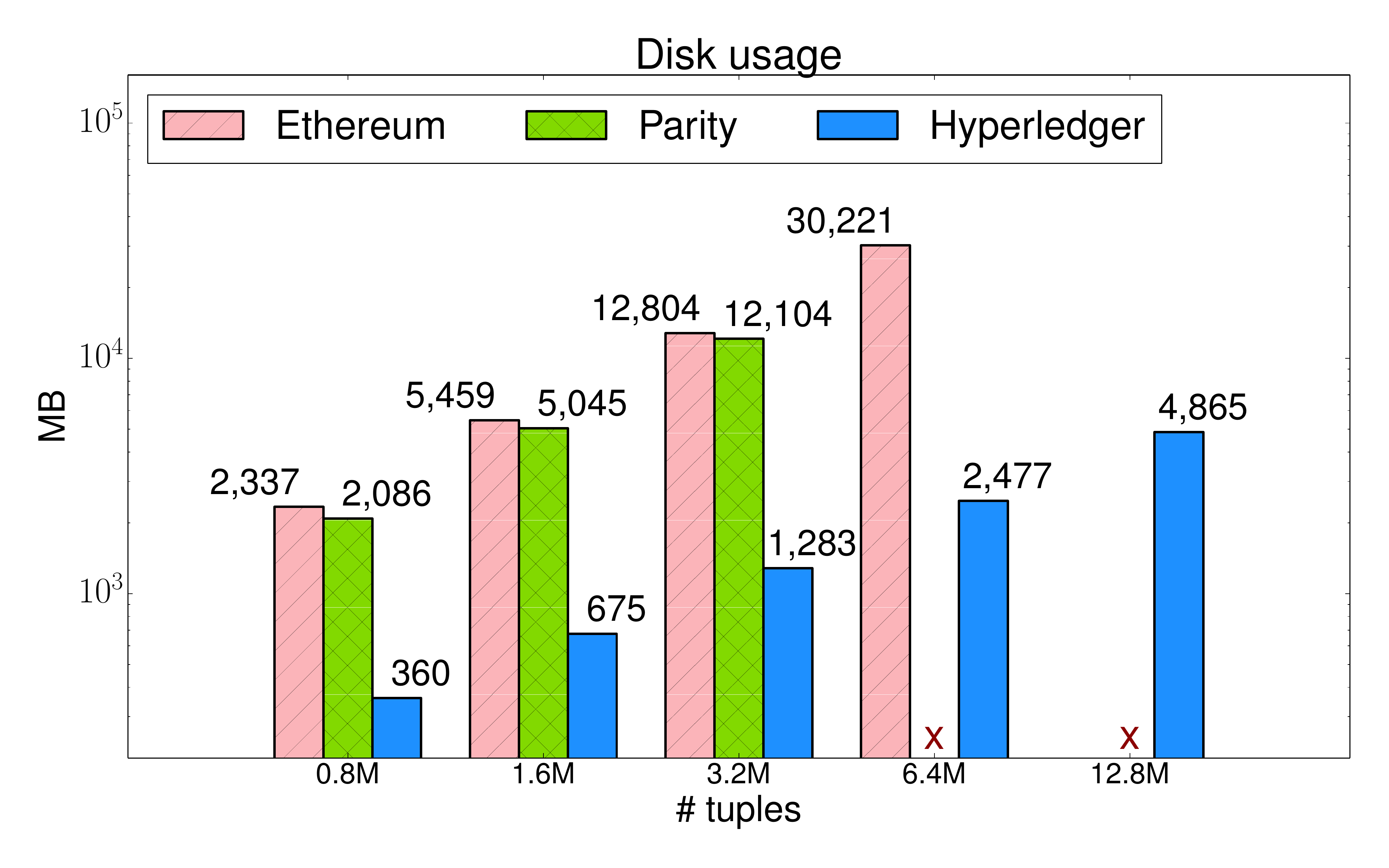}}
\caption{IOHeavy workload, {\bf `X'} indicates Out-of-Memory error.}
\label{fig:ioheavy}
\end{figure*}

\subsubsection*{Data model - IOHeavy}
\begin{figure*}
\centering
\subfloat[Analytics workload (Q1)]{\includegraphics[width=0.33\textwidth]{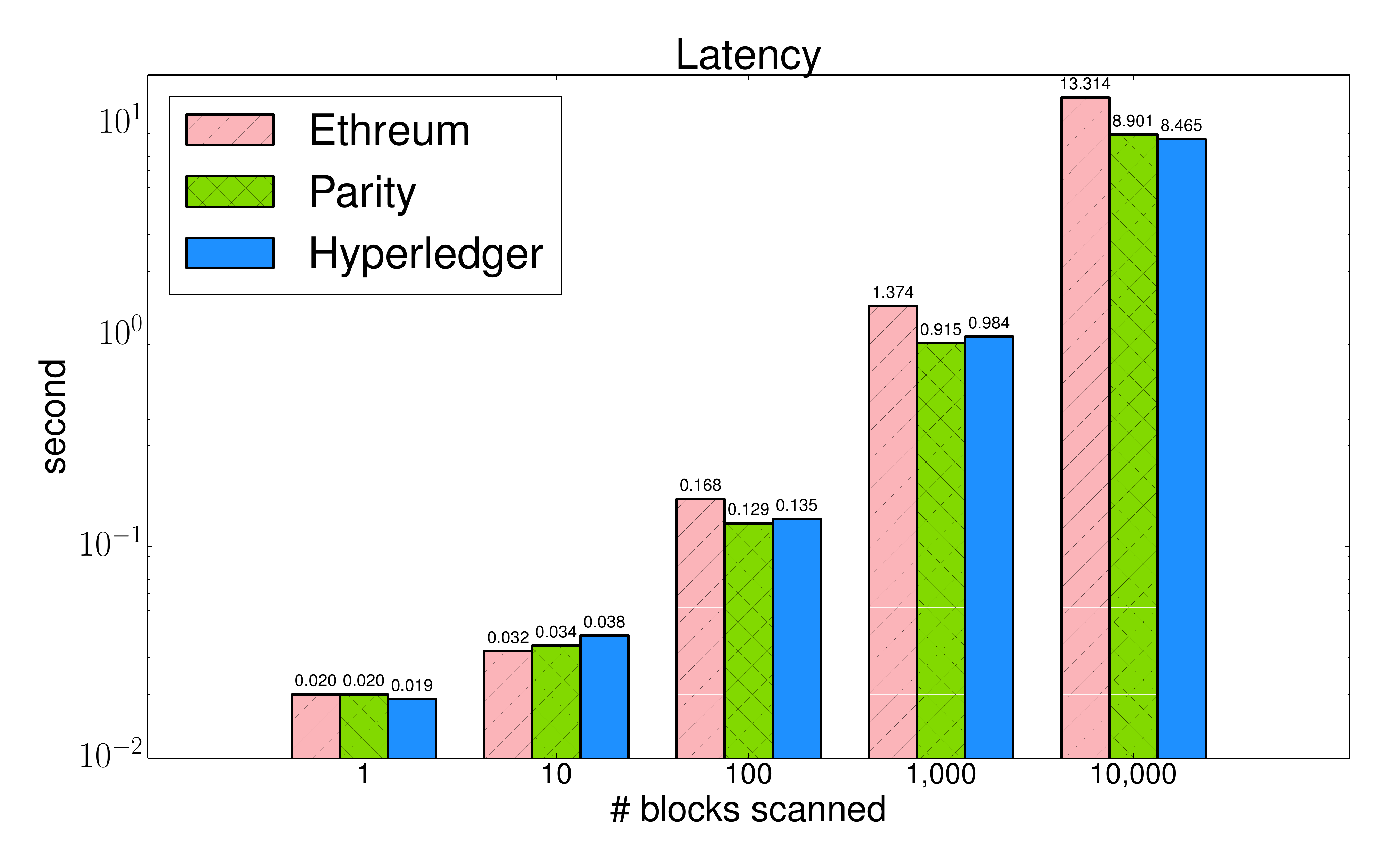}}
\subfloat[Analytics workload (Q2)]{\includegraphics[width=0.33\textwidth]{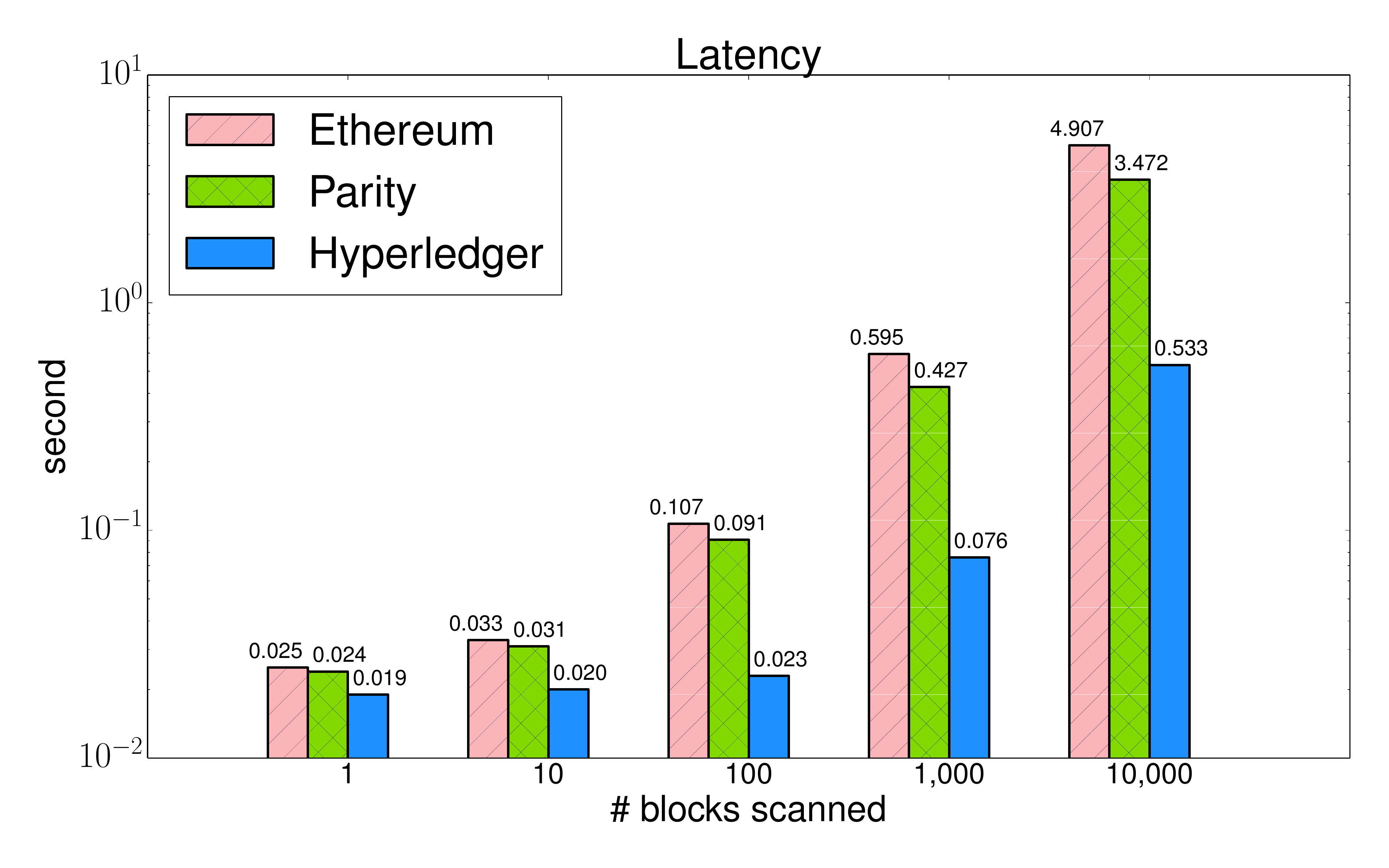}}
\subfloat[DoNothing workload]{\includegraphics[width=0.33\textwidth]{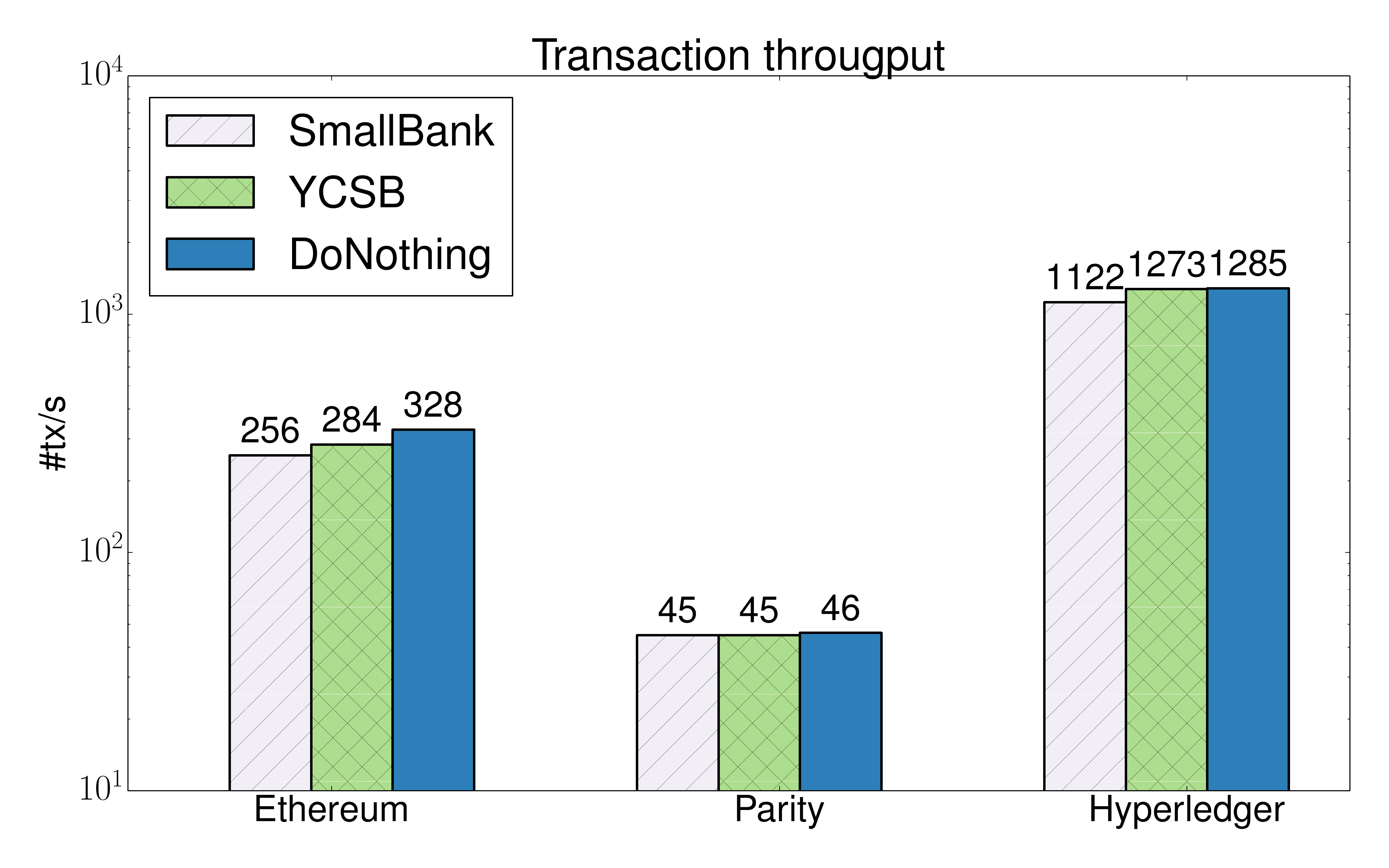}}
\caption{Analytics and DoNothing workloads.}
\label{fig:analytics}
\end{figure*}
We deployed the IOHeavy smart contract that performs a number of read and write operations of
key-value tuples. We used 20-byte keys and 100-byte values. Figure~\ref{fig:ioheavy} reports the throughput and disk
usage for these operations. Ethereum and Parity use the same data model and internal index structure, therefore they
incur similar space overheads. Both use an order of magnitude more storage space than Hyperledger which employs a simple
key-value data model. Parity holds all the state information in memory, so it has better I/O performance but fails to
handle large data (capped by over 3M states under our hardware settings). On the contrary, Ethereum only caches only
parts of the state in memory (using LRU for eviction policy), therefore it can handle more data than Parity at
the cost of throughput. Hyperledger leverages RocksDB to manage its states, which makes it more efficient at scale. 

\begin{figure*}
	\centering
	\subfloat[Write]{\includegraphics[width=0.33\textwidth]{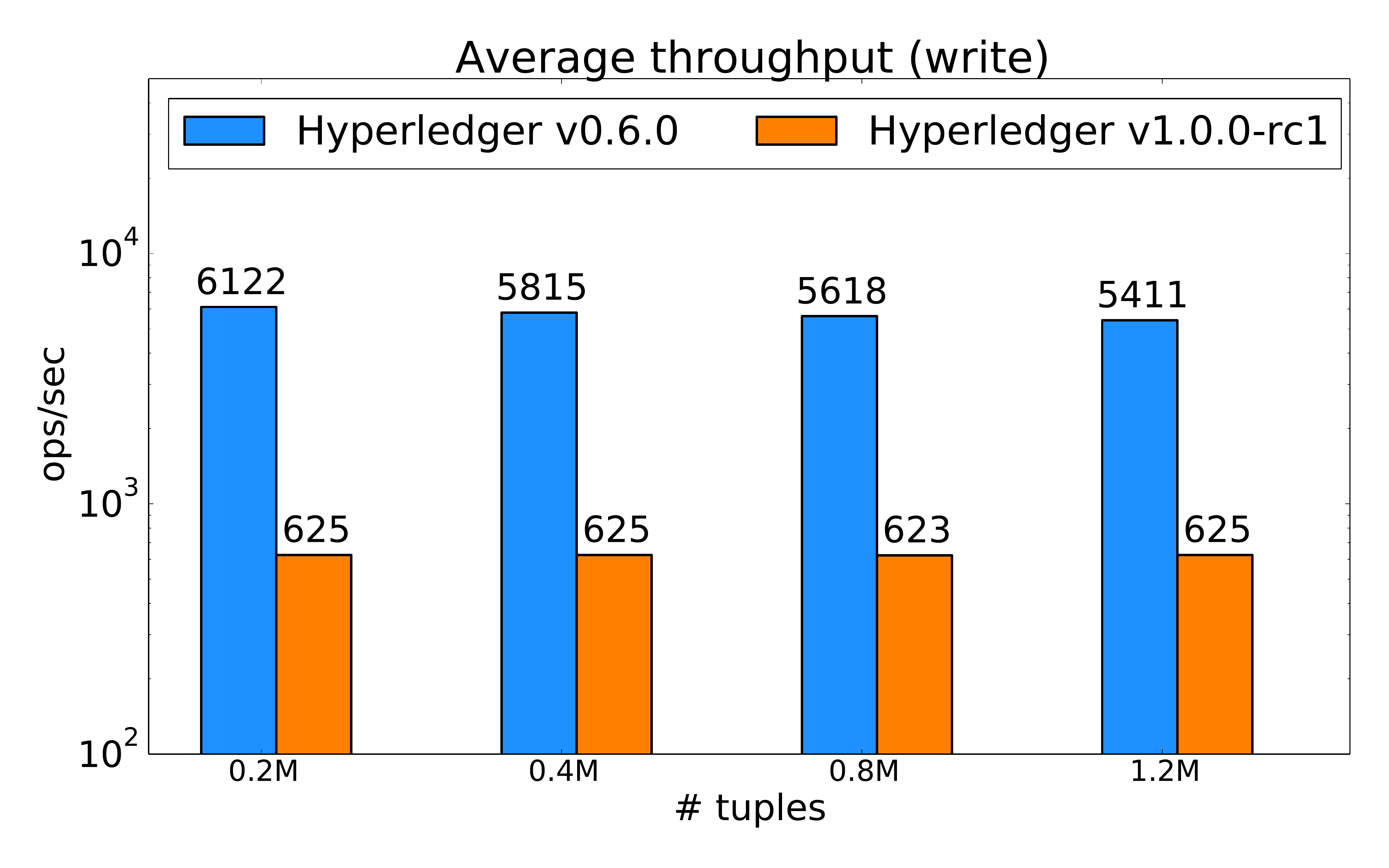}}
	\subfloat[Read]{\includegraphics[width=0.33\textwidth]{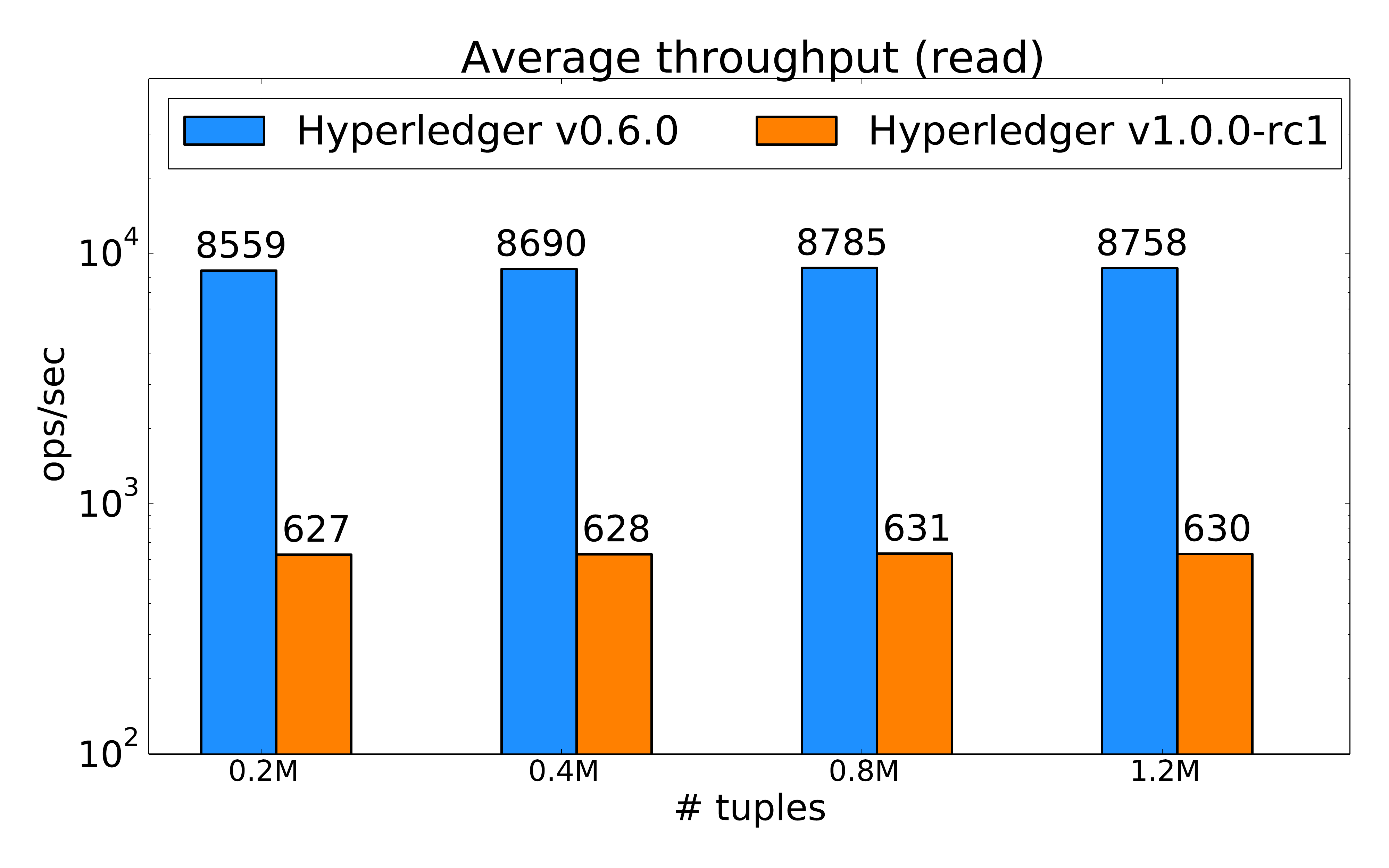}}
	\subfloat[Read]{\includegraphics[width=0.33\textwidth]{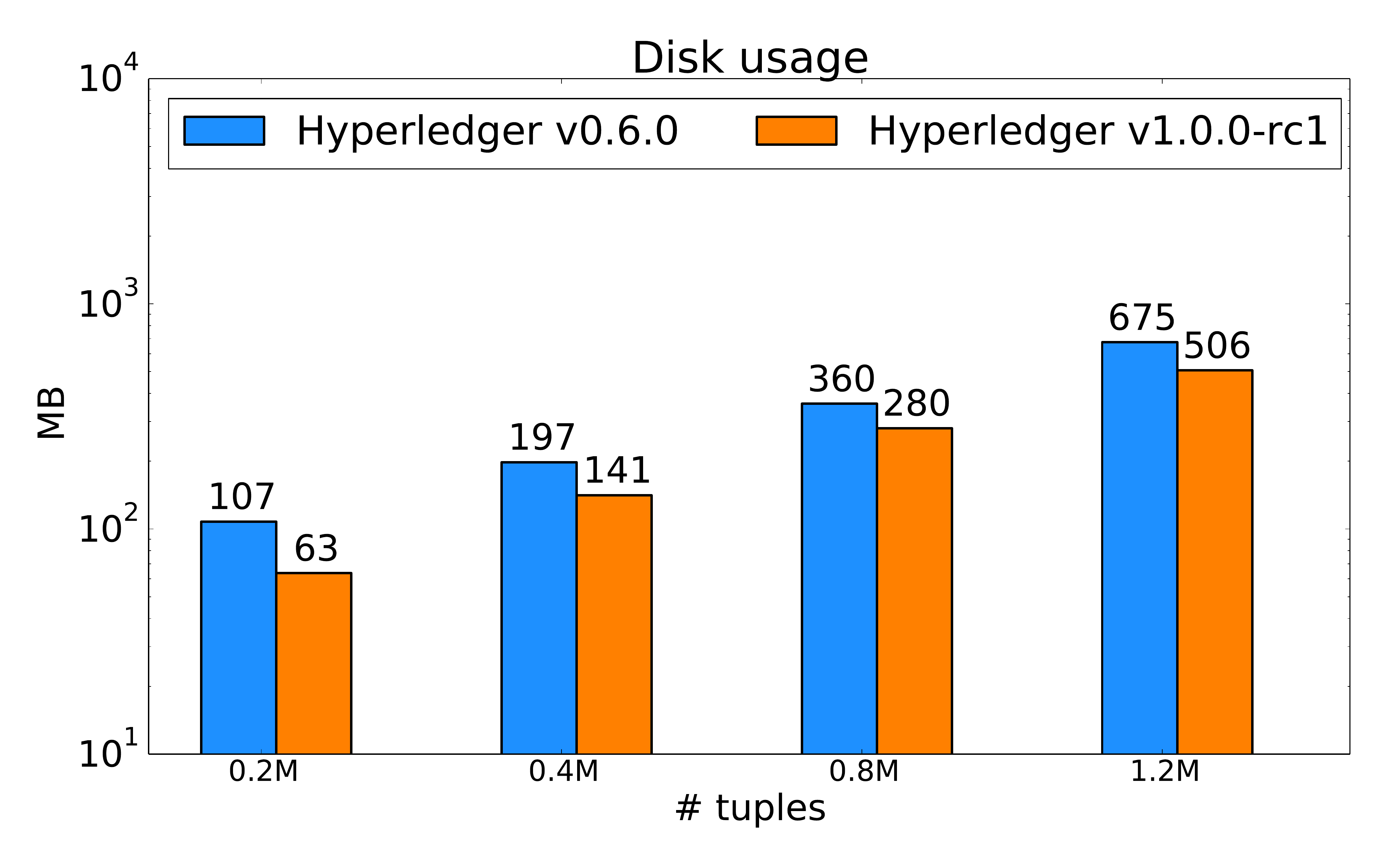}}
	\caption{Hyperledger v0.6.0 vs. v1.0.0 against IOHeavy workload.}
	\label{fig:ioheavy_n}
\end{figure*}

\vspace{0.5em}
\noindent \textbf{Hyperledger v1.0.}
We used the same IOHeavy smart contract to compare I/O performance of Hyperledger  version $v1.0$ with the
older version $v0.6$. As Figure~\ref{fig:ioheavy_n} illustrates, throughputs of $v1.0$ is an order of
magnitude worse than that of $v0.6$. Furthermore, $v1.0$ crashes with more than $0.8$M operations, reporting
exceptions about message oversizes. The significant gap can be attributed to the changes in the system
architecture from $v0.6$ to $v1.0$. In the former, the nodes take part in PBFT to confirm a block. In this
case, transactions in the IOHeavy workload incur no consensus overhead because there is only one node. In the
latter, a new service, {\em the orderer}, is introduced into the network to order transactions and provide the
consensus.  With this new service, transactions in the IOHeavy workload now need to communicate with the
orderer for them to be confirmed. More specifically, the nodes in $v1.0$ perform three more steps to finish a
transaction compared to $v0.6$. As communication overhead increases, the throughputs decrease. This result
suggests that replacing PBFT with a centralized service not only fails to protect the blockchain against
Byzantine failures, but it may also impair the overall performance. 

\subsubsection*{Data model - Analytics}
We implemented the analytics workload by initializing the three systems with over $120,000$ accounts
with a fixed balance. We then pre-loaded them with $100,000$ blocks, each contains 3 transactions on average. The
transaction transfers a value from one random account to another random account. Due to Parity's overheads in signing
transactions when there are many accounts, we considered transactions using only $1024$ accounts. We then executed the
two queries described in Section~\ref{subsec:workloads} and measured their latencies.  Figure~\ref{fig:analytics} shows
that the performance for Q1 is similar, whereas Q2 sees a significant gap between Hyperledger and the rest. We note that
the main bottleneck for both Q1 and Q2 is the number of network (RPC) requests sent by the client. For Q1, the client
sends the same number of requests to all systems, therefore their performance are similar. On the other hand, for Q2 the
client sends one RPC per block to Ethereum and Parity, but only one RPC to Hyperledger because of our customized smart
contract implementation. This saving in network roundtrip time translates to over $10$x improvement
in Q2 latency.  

\subsubsection{Consensus}
We deployed the DoNothing smart contract that accepts a transaction and returns immediately. We measured the throughput
of this workload and compare against that of YCSB and Smallbank. The differences compared to other workloads, shown in
Figure~\ref{fig:analytics}[c] is indicative of the cost of consensus protocol versus the rest of the software stack. In
particular, for Ethereum we observe $10\%$ increases in throughput as compared to YCSB, which means that execution of
the YCSB transaction accounts for the $10\%$ overhead. We observe no differences among these workloads in Parity,
because the bottleneck in Parity is due to transaction signing (even empty transactions still need to be signed), not
due to consensus or transaction execution.


%% file: db.tex
\section{Discussion}
\label{sec:db}
In this section, we first distill the lessons learned during the comparative studies of Ethereum, Parity and
Hyperledger. We then discuss how design principles from database systems could help improve blockchain
performance.  

\subsection{Lessons Learned From the Performance Study}
{\noindent\textbf{Understanding blockchain systems.} \name\ aims to facilitate better understanding of the
design and performance of different private blockchains. As more and more blockchain systems are being
proposed, each offering different sets of feature, \name's main value is that it narrows down the design space
into four distinct abstraction layers. The layers are derived from our taxonomy presented in Section~\ref{sec:tax}
which sufficiently captures the key and subtle characteristics of blockchain systems. By benchmarking these
layers, one can gain insights into the design trade-offs and performance bottlenecks. For example, using the
IOHeavy workload we identify that Parity trades performance for scalability by keeping states in memory. In
addition, the workload reveals potential performance issues with the latest version of Hyperledger.  Another
example is the Analytics workload that demonstrates trade-offs in the data models. In particular,
Hyperledger's simple key-value model means some analytical queries cannot be directly supported. However, it
enables optimization that helps answering analytical queries more efficiently.  Finally, we identify that the
bottleneck in Parity is not due to the consensus protocol, but due to the server's transaction signing. We
argue it is not easy to arrive at such insights without a systematic analysis framework.   

\noindent\textbf{Usability of blockchain.} Our experience in working with the three blockchain systems
confirms that in their current states, the blockchains are not yet ready for mass usage. Their designs and
codebases are still being refined constantly, and there are no other established applications beside
crypto-currency. Of the three systems, Ethereum is more mature both in terms of its codebase, user base and
developer community. Another usability issue we encountered is in porting smart contracts from one system to
another, due of their different programming models. This is likely to be exacerbated as more blockchain
platforms are being proposed~\cite{ripple,crypti,kadena,goodman2014tezos}.     

\noindent\textbf{Comparison to database systems.} The comparison against H-Store presented in the previous section
demonstrates that blockchains perform poorly at data processing tasks currently being handled by database systems.
Although databases are designed without security and tolerance to Byzantine failures, we remark that the gap
remains too big for blockchains to be disruptive to incumbent database systems. There are much for blockchains to learn
from databases in terms of high-performance data processing, which we discuss next. Nevertheless, there
are useful lessons that databases can take from the popularity and success of blockchains. Perhaps the most interesting
lesson is that there is a clear need for Byzantine tolerant data processing systems which can accommodate a large
number of users.  Distributed databases have diverged from P2P system designs by assuming non-Byzantine
failures~\cite{p2p}, but with the increasing availability of faster and more trustworthy hardware, this may be the right
time for the community to revise interest in high-performance, decentralized, P2P database systems.

\subsection{Bringing Database Designs into Blockchains}
\begin{figure*}
\centering
\includegraphics[width=0.85\textwidth]{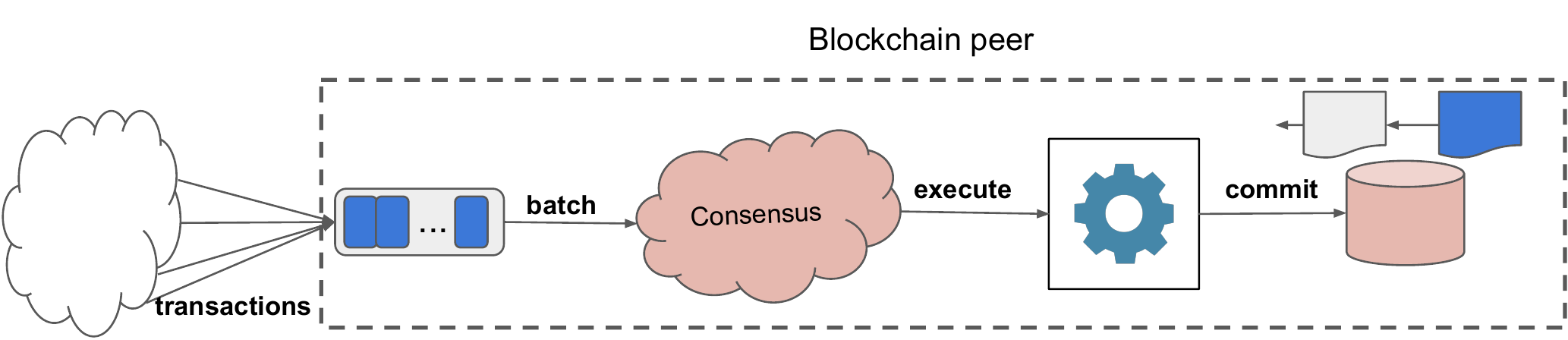}
\caption{The life of a blockchain transaction.}
\label{fig:lifecycle}
\end{figure*}
The challenges in scaling blockchain by optimizing the consensus protocols are being addressed in many recent
works~\cite{byzcoin,elastico}. Nevertheless, we demonstrated in our comparative study that there are other
performance bottlenecks beside consensus. Figure~\ref{fig:lifecycle} illustrates different stages that a
transaction goes through before it is considered committed to the blockchain. Each stage could be a potential
bottleneck and be subject to future optimizations. First, transactions are batch into a block. Next, the block
becomes input to the consensus protocol, and if selected by the protocol it is sent to the execution engine.
Finally, the engine executes the entire batch of transaction, creates new states and appends the block to the
chain\footnote{Note that the execution phase can be considered the last part of the consensus phase, because
during execution a node may detect conflicting transactions and abort the current consensus round.}.  We note
the striking resemblance with the flow of a transaction in a distributed database.  In fact, the only
difference being the consensus protocol: databases use two-phase commit or Paxos, whereas blockchains use
Byzantine tolerant protocols or variants thereof. Given the similarity, we propose four approaches
inspired by design principles in databases to improve blockchain performance. 

\vspace{0.25cm}
\noindent\textbf{Decoupling the layers and optimize them individually.} One possible direction is to decouple
storage, execution engine and consensus layer from each other, then optimize and scale them independently.
Tezos and Corda, for examples, have decoupled the consensus layer by outsourcing it to separate parties. The
data model layer could also be decoupled. For instance, current systems employ generic key-value storage,
which may not be best suited to the unique data structure and operations in blockchain. UStore~\cite{ustore}
demonstrates that a storage designed around the blockchain data structure is able to achieve better
performance than existing implementations. Most importantly, we observe that current data models in Ethereum
and Hypereldger are not ideal for answering analytics queries. In particular, both do not support fine-grained
versioning at the transaction level: it is not possible to immediately query previous versions of a state
except for querying at block level and processing the state of each block. Implementing a new data model is
less complex when the storage layer is decoupled from the rest of the blockchain stack. 

\vspace{0.25cm}
\noindent {\bf Embracing new hardware primitives.} Many data processing systems are taking advantage of new
hardware to boost their performance~\cite{in-memory,zhang15,farm}. For blockchain, using trusted hardware, the
underlying Byzantine fault tolerance protocols can be modified to incur fewer network messages~\cite{a2m}.
With trusted hardware, the blockchain can tolerate more failures, and because of fewer network messages it can
scale better.  Systems like Parity and Ethereum can take advantage of multi-core CPUs and large memory to
improve contract execution and I/O performance.

\vspace{0.25cm}
\noindent\textbf{Sharding.} Blockchain is essentially a replicated state machine system, in which each
node maintains the same data. As such, blockchains are fundamentally different to database systems such as
H-Store in which the data is partitioned (or sharded) across the nodes. Sharding helps reduce the computation
cost and can make transaction processing faster. The main challenge with sharding is to ensure consistency
among multiple shards. However, existing consistency protocols used in database systems do not work under
Byzantine failure. Nevertheless, their designs can offer insights into realizing a more scalable sharding
protocol for blockchain. Corda partitions the ledger into sub-ledgers belonging to smaller group, thus
avoiding the need to broadcast transactions to the entire network. However, it still depends on an external,
centralized component for consensus between sub-ledgers. Recent work~\cite{elastico} has demonstrated
the feasibility of sharding the consensus protocol, making important steps towards partitioning the entire
blockchain.

\vspace{0.5cm}
\noindent\textbf{Supporting declarative language.} Having a set of high-level operations that can be composed in
a declarative manner makes it easy to define complex smart contracts. It also opens up opportunities for
low-level optimizations that speed up contract execution. Hawk~\cite{kosba16} is a recent example that
applies compiler techniques to hide complexity of implementing privacy preserving smart contracts. However,
it aims to strengthen security of the smart contracts rather than improve their performance.

%% file: conclusion.tex
\section{Conclusion}
\label{sec:conclusion}
In this paper, we have conducted a comprehensive survey on blockchain technologies. We laid out four
underpinning concepts behind blockchains and analyzed the state of the art using these concepts. We presented
our benchmarking framework, \name\, which is designed to evaluate performance of blockchains as data
processing platforms. Finally, we discussed four potential research directions, inspired by database design
principles, for improving blockhchain performance. We hope that the survey and benchmarking framework would
serve to guide the design and implementation of future blockchain systems that are not only secure, but 
scalable and usable in the real world.

 